\documentclass[twocolumn]{aastex7}

\usepackage{xspace}
\usepackage{amsmath}
\usepackage{mathrsfs}
\usepackage{hyperref}
\newcommand{\jwst}{{\it JWST}\xspace}

\newcommand{\E}{$E\left(B-V\right)$}

\newcommand{\R}{$f_\mathrm{em}/f_\mathrm{abs}$}

\newcommand{\ggi}{SN~2024ggi}

\begin{document}

\title{Probing the 3D Structures of Supernovae through IR Signatures of CO and SiO}

\newcommand{\PSI}{\affiliation{Planetary Science Institute, 1700 East Fort
  Lowell Road, Suite 106,Tucson, AZ 85719-2395 USA}}
\newcommand{\HS}{\affiliation{Hamburger Sternwarte, Gojenbergsweg 112, 21029 Hamburg, Germany}}
\newcommand{\IFA}{\affiliation{Institute for Astronomy, University of Hawai’i at Manoa, 2680 Woodlawn Dr., Hawai’i, HI 96822, USA}}
\newcommand{\VT}{\affiliation{Department of Physics, Virginia Tech,
    850 West Campus  Drive, Blacksburg VA, 24061, USA}}
\newcommand{\GRFP}{\altaffiliation{National Science Foundation Graduate Research Fellow}}

\newcommand{\FINESST}{\altaffiliation{NASA FINESST Future Investigator}}
\newcommand{\NHFPE}{\altaffiliation{NHFP Einstein Fellow}}
\newcommand{\UIUC}{\affiliation{Department of Astronomy, University of Illinois Urbana-Champaign, 1002 West Green Street, Urbana, IL 61801, USA}}
\newcommand{\NSFSIMS}{\affiliation{NSF-Simons AI Institute for the Sky (SkAI), 172 E. Chestnut St., Chicago, IL 60611, USA}}
\newcommand{\STSci}{\affiliation{Space Telescope Science Institute, 3700 San Martin Drive, Baltimore, MD 21218-2410, USA}}
\newcommand{\FSU}{\affiliation{Department of Physics, Florida State
    University, Tallahassee, FL 32306, USA}}
\newcommand{\Carnegie}{\affiliation{Observatories of the Carnegie
    Institution for Science, 813 Santa Barbara St., Pasadena, CA 91101, USA}}
\newcommand{\MSU}{\affiliation{Department of Physics \& Astronomy,
    Michigan State University, East Lansing, MI, USA}}
\newcommand{\TAMU}{\affiliation{George P. and Cynthia Woods Mitchell
    Institute for Fundamental Physics and Astronomy,
    Department of Physics and Astronomy, Texas 
             A\&M University, College Station, TX 77843, USA}}
\newcommand{\IALP}{\affiliation{Instituto de Astrof\'isica de La Plata
    (IALP), CONICET, Paseo del Bosque S/N, B1900FWA La Plata, Argentina}}
\newcommand{\LaPlata}{\affiliation{Facultad de Ciencias Astron\'omicas
    y Geof\'isicas Universidad Nacional de La Plata, Paseo del Bosque,
    B1900FWA, La Plata, Argentina}}
\newcommand{\WPI}{\affiliation{Kavli Institute for the Physics and
    Mathematics of the Universe (WPI), The University of Tokyo,
    Kashiwa, 277-8583 Chiba, Japan}} 

\newcommand{\ICE}{\affiliation{Institute of Space Sciences (ICE,
    CSIC), Campus UAB, Carrer de Can Magrans, s/n, E-08193 Barcelona, Spain}}

\newcommand{\IEEC}{\affiliation{Institut d’Estudis Espacials de
    Catalunya (IEEC), E-08034  Barcelona, Spain}} 

\newcommand{\LCO}{\affiliation{Las Campanas Observatory, Carnegie
    Observatories, Casilla 601, La Serena, Chile}} 

\newcommand{\Aarhus}{\affiliation{Department of Physics and Astronomy,
    Aarhus University, Ny  Munkegade 120, DK-8000 Aarhus C, Denmark.}} 

\newcommand{\OU}{\affiliation{Homer L.~Dodge Department of Physics and
  Astronomy, University of Oklahoma, 440 W. Brooks, Rm 100, Norman, OK
  73019-2061}}  

\newcommand{\UCSC}{\affiliation{Department of Astronomy and Astrophysics,
  University of California, Santa Cruz, CA 95064, USA}} 
\newcommand{\Melbourne}{\affiliation{School of Physics, The University of
  Melbourne, VIC 3010, Australia}}

\newcommand{\LPNHE}{\affiliation{LPNHE, (CNRS/IN2P3, Sorbonne
  Universit\'e, Universit\'e Paris Cit\'e), Laboratoire de Physique
  Nucl\'eaire et de Hautes \'Energies, 75005, Paris, France}}

\newcommand{\Princeton}{\affiliation{Princeton University, 4 Ivy Lane,
    Princeton, NJ 08544, USA}}

\newcommand{\Berkeley}{\affiliation{Department of Astronomy,
    University of California, Berkeley, CA 94720-3411, USA}}

\newcommand{\Tsinghua}{\affiliation{Physics Department, Tsinghua
    University, Beijing, 100084, China}}

\newcommand{\Thailand}{\affiliation{National Astronomical Research
    Institute of Thailand, 260 Moo 4, Donkaew, Maerim, Chiang Mai,
    50180, Thailand}}

\newcommand{\UVA}{\affiliation{Department of Astronomy, University of
    Virginia, 530 McCormick Rd, Charlottesville, VA 22904, USA}}

\newcommand{\LJMU}{\affiliation{Astrophysics Research Institute,
    Liverpool John Moores University, 146 Brownlow Hill, Liverpool L3
    5RF, UK}}

\newcommand{\MPIA}{\affiliation{Max-Planck-Institut f\"ur Astrophysik,
    Karl-Schwarzschild Stra{\ss}e 1, 85748 Garching, Germany}}

\newcommand{\JHU}{\affiliation{Physics and Astronomy Department,
    Johns Hopkins University, Baltimore, MD 21218, USA}}

\newcommand{\OSU}{\affiliation{Department of Astronomy, The Ohio State
    University, Columbus, OH, USA}}

\newcommand{\CCAP}{\affiliation{Center for Cosmology and Astroparticle
    Physics, The Ohio State University, Columbus, OH, USA}}

\newcommand{\MIT}{\affiliation{Department of Physics and Kavli Institute for Astrophysics and Space Research, Massachusetts Institute of Technology, 77 Massachusetts Avenue, Cambridge, MA 02139, USA}}

\newcommand{\nextinstitute}{\affiliation{Put the institute of the new author here}}

\author[0000-0001-5888-2542]{T.~Mera}
\email{tycomera@gmail.com}
\FSU

\author[0000-0002-4338-6586]{P.~Hoeflich}
\email{phoeflich77@gmail.com}
\FSU

% \author[0000-0002-7566-6080]{J. M. DerKacy}
% \email{jderkacy@stsci.edu}
% \STSci

% \author[0000-0002-4338-6586]{N.~Morrell}
% \email{nmorrell@carnegiescience.edu}
% \LCO

% \author[0000-0001-9668-2920]{J.~T.~Hinkle}
% \NHFPE
% \UIUC
% \NSFSIMS
% \IFA
% \email{jhinkle6@hawaii.edu}

% % \author[0000-0001-6272-5507]{P. J.~Brown}
% % \email{grbpeter@yahoo.com}
% % \TAMU

% \author[[0000-0001-6876-8284]{P. A. Mazzali}
% \email{P.Mazzali@ljmu.ac.uk}
% \LJMU
% \MPIA

\author[0000-0003-4625-6629]{C.~R.~Burns}
\email{cburns@carnegiescience.edu}
\Carnegie

\author[0000-0002-5221-7557]{C. Ashall}
\email{cashall@hawaii.edu}
\IFA

\author[0000-0001-7186-105X]{K. Medler}
\email{kyle.medler@sky.com}
\IFA

\author[0009-0001-9148-8421]{E.~Fereidouni}
\email{ef22g@fsu.edu}
\FSU

\author[0000-0003-3953-9532]{W.~B.~Hoogendam}
\GRFP
\IFA
\email{willemh@hawaii.edu} 

\author[0000-0002-9301-5302]{M.~Shahbandeh}
\email{mshahbandeh@stsci.edu}
\STSci

\author[0000-0001-6107-0887]{S. Shiber}
\FSU
\email{sshiber1@lsu.edu}

\author[0000-0002-7305-8321]{C.~M.~Pfeffer}
\GRFP
\IFA
\email{cpfeffer@hawaii.edu}

\author[0000-0001-5393-1608]{E.~Baron}
\email{ebaron@psi.edu}
\PSI
\HS

\author[0000-0002-3900-1452]{J. Lu}
\email{lujingeve158@gmail.com}
\MSU

\author[0000-0002-4338-6586]{N.~Morrell}
\email{nmorrell@carnegiescience.edu}
\LCO

\author[0000-0003-1039-2928]{E.~Y.~Hsiao}
\email{yichi.hsiao@gmail.com}
\FSU

\author[0000-0003-2734-0796]{M.~M.~Phillips}
\email{mmp@lco.cl}
\LCO

\submitjournal{ApJ}

\received{\today}
\revised{\today}
\accepted{\today}

\begin{abstract}
We present a new public-domain MOlecular Fitting Analysis Tool (MOFAT) designed to probe molecule-forming regions in supernovae (SNe) through analysis of molecular features in the near- and mid-infrared. MOFAT employs a novel data-driven approach to explore the physical properties of these regions using time-independent radiative transfer simulations that include multidimensional, clump-like structures, constrained by high-precision observations. Such structures are required to reproduce the flux ratio between fundamental and overtone bands, overcoming limitations of traditional one-zone forward-modeling, such as optical-depth effects and initial configurations. Our approach enables spectral fits that can reconstruct overall abundances and temperatures and determine parameterized small-scale structures associated with physical instabilities.
We systematically study the relationship between physical parameters and the profiles of CO and SiO, showing that free parameters are constrained, while detection of small-scale structure requires optically thick bands. As a demonstration, MOFAT is applied to \ggi\ at +285 and +385 days post-explosion. We find that CO formation triggers SiO formation in the inner layers of the CO-rich region previously studied. The inner edge of the SiO-emitting region recedes from velocities of $v_1 \approx 1{,}500$ to $1{,}000~\mathrm{km~s^{-1}}$, indicating continued SiO formation.
The SiO mass decreases from $\sim(2-6)\times10^{-3}~M_\odot$ by roughly an order of magnitude, suggesting ongoing evaporation. SiO features indicate clumping, but most of the flux originates from optically thin regions. SiO contributes negligibly to cooling, and we find no evidence for dust formation. Finally, we discuss observational strategies to trace the evolution of molecule formation and its connection to dust formation.

\end{abstract}

%%%%%%%%%%%%%%%%%%%%%%%%%%%%%%%%%%%%%%%%%%%%%%%%%%

%%%%%%%%%%%%%%%%% BODY OF PAPER %%%%%%%%%%%%%%%%%%

\section{Introduction} \label{sec:intro}

Massive stars ($\approx 8-30$ M$_{\odot}$) end their lives as core-collapse supernovae (CC~SNe) and enrich the interstellar medium with iron-group and $\alpha$ elements \citep{woosley_evolution_1995,janka_theory_2007,smartt_progenitors_2009}. It has also been suggested that CC~SNe significantly contribute to observed dust masses in the early universe, but theory has been at odds with observations \citep{wooden_airborne_1993,maiolino_supernova_2004,dwek_evolution_2007,kotak_dust_2009,matsuura_dust_2017,sarangi_dust_2018,shahbandeh_jwst_2023,shahbandeh_jwstmiri_2025}. 

Prior to dust formation, CC~SNe are expected to synthesize 
Carbon Monoxide (CO) and Silicon Monoxide (SiO). The first overtone and fundamental ro-vibrational bands of these molecules radiate strongly in near- and mid-infrared (NIR and MIR) wavelengths, which act as coolants to catalyze the growth of dust grains \citep{liu_carbon_1992,liu_silicon_1994,cherchneff_chemistry_2009,sluder_molecular_2018,liljegren_carbon_2020,liljegren_molecular_2023,mcleod_carbon_2024,cherchneff_revisiting_2025}. Despite the importance of this cooling, 
the distribution of dust nucleation sites through CO and SiO need to be studied in detail. This task has proven difficult because both observational and theoretical evidence suggest that the ejecta of CC~SNe are structured and mixed on various instability scales, making direct comparison between the two computationally complex 
\citep{fryxell_instabilities_1991,khokhlov_jet-induced_1999,couch_aspherical_2009,wallstrom_co_2013,abellan_very_2017,dessart_impact_2018,law_three-dimensional_2020,ono_matter_2020,priestley_constraining_2020,rho_near-infrared_2021,soker_role_2022,burrows_physical_2024,temim_dissecting_2024,vartanyan_3d_2025}.

\begin{figure*}[t]
    \centering
    \includegraphics[width=0.8\linewidth]{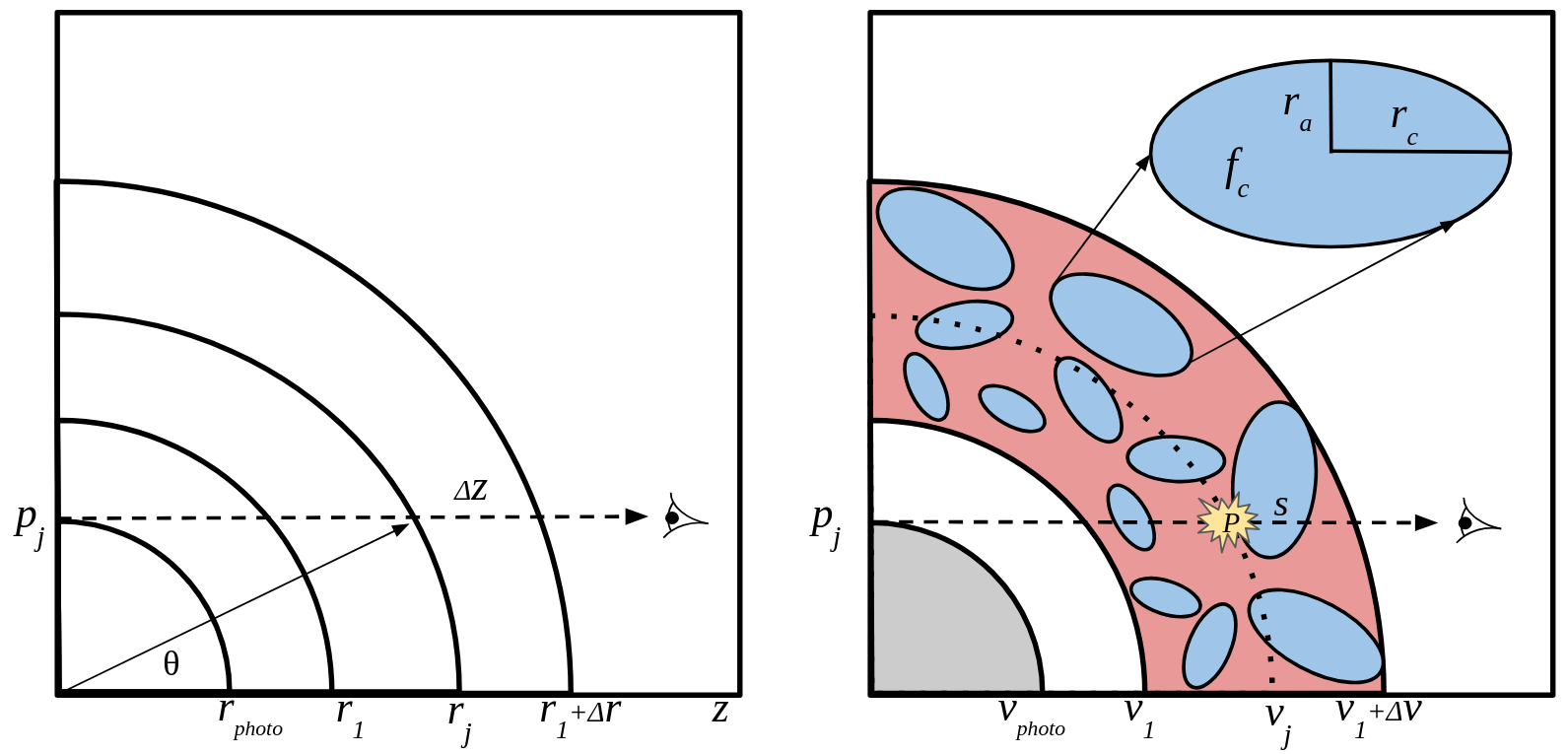}
    \caption{Left: The coordinate systems of $(r,\theta)$ and $(p,z)$. $r_{photo}$ is the photosphere, $r_1$ and $r_1+\Delta r$ are the inner and outer radius of the molecular shell, respectively, and $r_j$ is some arbitrary radius. These locations coincide with the equivalent velocities in the right figure. Right: A cartoon graphic of how MOFAT works. The molecular clumps have semi-axes of $r_c$ and $r_a$ with a density enhancement factor of $f_c$. When we integrate along $p_j$, the ray will have a certain probability ($P$, yellow star) of hitting a clump at $v_j$. The effects of the clump are then included, which are based on the chord length $s$. See Appendix \ref{sec:global_rad} for variable descriptions and implementation.}
    \label{fig:mofat}
\end{figure*}

Current time-dependent forward modeling employs spherical models. This class of models successfully explains the timing of molecule formation. As a result, such models are able to fit the optically thin portions of the CO overtone and fundamental bands, which can be observed from the ground and with the \textit{Spitzer Space Telescope} \citep{hoflich_spectral_1988, spyromilio_carbon_1988, liu_carbon_1992, gerardy_detection_2000, gerardy_carbon_2002}.

With the \textit{James Webb Space Telescope} (\jwst), optically thick profiles of the fundamental band and optically thin profiles of the overtone bands are routinely observed. However, spherical forward models fail to reproduce the absolute flux ratio between these bands by factors of 2–5 \citep{mcleod_carbon_2024}. An obvious solution is the inclusion of small-scale 3D structures, which reduces the effective optical depth, but introduces a large parameter space that must be tested.

With the rapidly increasing NIR and MIR datasets from \jwst, we are left with the question: How can we effectively and efficiently model the 3D structure of the molecular region in CC~SNe to reproduce CO and SiO observations of the overtone and fundamental bands at any epoch?

Instead of optimizing forward modeling methods to continuously and randomly test different explosion instabilities against observations, we present an inverse data-driven approach. Namely, we utilize observations to drive full radiation transport simulations that include 3D morphology effects to reproduce spectra at a given time. We aim to describe the structures of the molecular region as it is seen, giving us a powerful tool to probe the molecular region in detail. 

We organize our manuscript as follows: Section \ref{sec:motivation} provides context for our motivation, Section \ref{sec:model} introduces our MOlecular Fitting Analysis Tool (MOFAT), Section \ref{sec:results} discusses the stability of MOFATs solutions for the CO and  SiO features of \ggi, Section \ref{sec:dis} discusses the implications of MOFAT, and Section \ref{sec:con} concludes our study.

\section{Motivation}\label{sec:motivation}

When describing the molecular region for CC~SNe, the physical conditions of the progenitor before collapse, the explosion mechanism, and the source and scales of instabilities during the explosive phase play a major part in the resulting structure. This stratification will play a major role in molecular formation because, for a given CC~SN, it is unclear when, where, or how much CO and SiO will form. As an example, extensive analysis of SN~1987A has shown that explosive and stellar burning layers may be significantly mixed and sufficiently clumped, which has led to CO being observed at velocities of $\sim3000$~$km~s^{-1}$ and H lines at $\sim700$~$km~s^{-1}$ 
\citep{hoflich_spectral_1988,spyromilio_carbon_1988,hanuschik_absolute_1988,petuchowski_co_1989,arnett_supernova_1989,spyromilio_spectral_1990,li_iron_1993,basko_nickel_1994}

The evolution of massive stars influences the chemical layering prior to collapse. Stellar evolutionary codes successfully simulate the life of these objects through sequences of nuclear burning stages 
\citep{woosley_evolution_2002,ekstrom_effects_2008,heger_nucleosynthesis_2010,chieffi_pre-supernova_2013,paxton_modules_2015,limongi_presupernova_2018,paxton_modules_2019}, however, uncertainties both physically and computationally affect their evolution and resulting stratification \citep{boccioli_neutrino_2025}. In particular, mass-loss 
\citep{langer_presupernova_2012,smith_mass_2014}, mixing from convective and rotation processes 
 \citep{meynet_stellar_1997,arnett_beyond_2015,maederPhysicsFormationEvolution2009,joyce_review_2023}, magnetic fields \citep{maeder_stellar_2003,maeder_stellar_2005,varma_3d_2021,hirschi_stellar_2024}, and binary and multiple systems 
\citep{laplace_different_2021,farmer_nucleosynthesis_2023,marchant_evolution_2024} will strongly modify the chemical layout prior to collapse. 

It is commonly accepted that the explosion of successful CC~SNe undergoes a delayed neutrino heating mechanism to revive a stalled shock front \citep{bethe_revival_1985,boccioli_neutrino_2025}. 3D simulations have been shown to self consistently explode but are computationally intense, limiting parametric studies 
\citep{muller_new_2012,takiwaki_three-dimensional_2016,janka_physics_2016,burrows_overarching_2020,nakamura_three-dimensional_2022,vartanyan_3d_2025}. This makes 1D and 2D simulations practical, but they pose an explodability problem and typically require an artificial explosion method (e.g. a thermal or kinetic bomb) or additional effects due to neutrino transport 
\citep{woosley_evolution_1995,heger_nucleosynthesis_2010,chieffi_pre-supernova_2013,limongi_hydrodynamical_2020,couch_simulating_2020,wang_essential_2022,boccioli_explosion_2023,maltsev_explodability_2025}. In the case of the artificial explosion method, the total explosion energy can be tuned to observables; however, this adds another level of parameterization, which further complicates the resulting molecular region due to dependence on the location of the `mass cut' and on the amount of
energy deposition via explosive nucleosynthesis 
\citep{woosley_evolution_1995,thielemann_core-collapse_1996,young_uncertainties_2007,tominaga_supernova_2007,heger_nucleosynthesis_2010,jerkstrand_constraints_2015,wanajo_nucleosynthesis_2018,sawada_nucleosynthesis_2019,reichert_nucleosynthesis_2021,roberti_ensuremathgamma-process_2024,west_impact_2025,limongi_chemical_2025}. For neutrino-driven explosions, neutrino luminosity has been taken as a semi-free parameter in some studies 
\citep{ugliano_progenitor-explosion_2012,perego_pushing_2015,sukhbold_core-collapse_2016,ertl_two-parameter_2016,curtis_pushing_2019,ebinger_pushing_2020}, but recently, 1D and 2D simulations that include turbulent effects seen in 3D simulations are more self-consistent and have shown reliance on the 
Si/O interface, making the molecular forming layers a region where turbulent effects begin to play a heightened role 
\citep{couch_simulating_2020,wang_essential_2022,kumar_explode_2024,boccioli_neutrino_2025,boccioli_quantifying_2025}.

\begin{figure*}[t]
    \centering
    \includegraphics[width=0.65\linewidth]{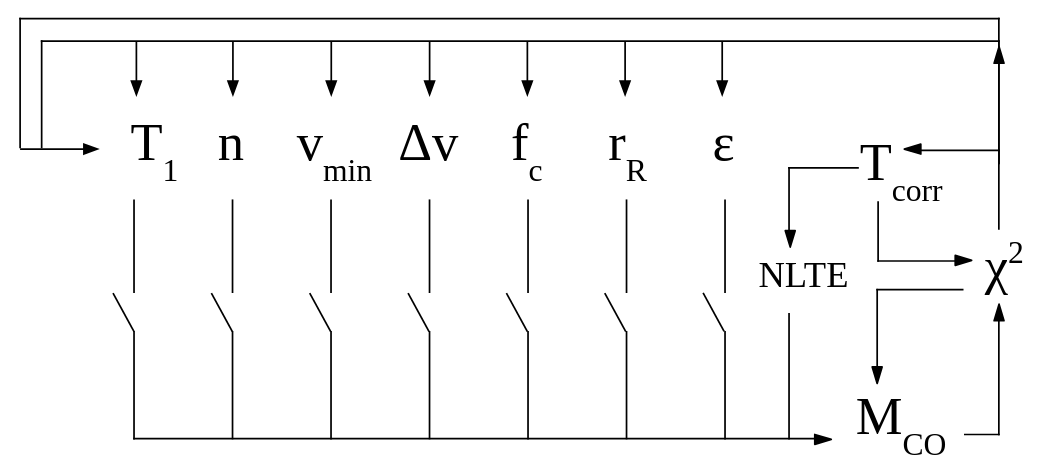}
    \caption{A schematic showing the iteration scheme used in MOFAT. The outermost loop loops through blocks, where each block closes particular switches for parameters it wants to test. Each selected parameter is then looped through, where temperature, NLTE, and total mass effects are iterated until the $\chi^2$ statistic reaches a convergence criterion or a maximum number of iterations. This block can then be looped through, where between loops the parameters region is centered on the best value and subsequently shrunk by 10\%. See Appendix \ref{sec:iteration} for a more complete description.}
    \label{fig:mofat_iteration}
\end{figure*}

It is well established that multidimensional effects are necessary to get a successful explosion 
\citep{janka_physics_2016,muller_hydrodynamics_2020,janka_long-term_2025}. Turbulent motions in the late stages of a star's life may act as seeds for the multidimensional effects that trigger a successful explosion (e.g., jets, Standing Accretion Shock Instabilities (SASI), or neutrino-driven turbulent convection) 
\citep{khokhlov_jet-induced_1999,wheeler_asymmetric_2002,blondin_stability_2003,foglizzo_neutrino-driven_2006,couch_aspherical_2009,soker_applying_2010,hanke_is_2012,murphy_dominance_2013,nagakura_importance_2019,fields_three-dimensional_2021}. These instabilities will then grow as Rayleigh-Taylor (RT) or Richtmyer-Meshkov (RM) instabilities as the shock propagates outwards through the progenitor 
\citep{fryxell_instabilities_1991,kifonidis_non-spherical_2003,utrobin_three-dimensional_2019,burrows_three-dimensional_2019}. The scaling of these instabilities is situationally and directionally dependent, adding to the complex nature that might exist in the Si/O/C layers of CC~SNe.

Given the plethora of uncertainties that surround successful CC~SNe explosions and the variations in instability and mixing scales that may arise in the molecular forming layers, forward modeling becomes exceedingly difficult when attempting to explain diatomic molecular features. In particular, sufficient computational resolution is required to capture these instabilities in synthetic spectra. These limitations have made the reproduction of the flux ratio between the overtone and fundamental band a challenge. Moreover, prior to \jwst, these methods, along with `quick-fit' techniques, have been instrumental in explaining the CO first overtone, helping decipher the total molecular mass, the temperature at the narrow decoupling region of the overtone, and the magnitude of mixing 
\citep{sharp_molecular_1990,liu_carbon_1992,liu_silicon_1994,gerardy_detection_2000,gerardy_carbon_2002,kotak_early-time_2005,das_detection_2009,hunter_extensive_2009,drout_double-peaked_2016,banerjee_early_2018,rho_near-infrared_2018,rho_near-infrared_2021}. 

With \jwst, we can now consistently observe the fundamental bands of CO and SiO, whose opacities are about two orders of magnitude larger than their overtones. This separation further complicates the molecular forming region because the temperature profile is now dictated by optical depth effects, which are directly tied to the level of clumping and total molecular mass, leading to variations in cooling, heating, formation, and destruction rates 
\citep{sluder_molecular_2018,liljegren_carbon_2020}.

With all of these obstacles in place, radiation transport simulations based on observations may provide an improved understanding on the molecular forming region at a given epoch.

% With all of these obstacles in place, it is more applicable to let nature tell us what path it chose by using observations to describe the molecular forming region. To do this, we will let observations drive
% % use observations by driving 
% radiation transport simulations at a given epoch.

\section{MOFAT}\label{sec:model}

The goal of MOFAT is to describe the molecular region of CC~SNe using observations.
% as observations would dictate. 
From Section \ref{sec:motivation}, MOFAT needs the ability to capture the resulting Si/O/C chemical stratification that includes cooling effects and clump-like structures. Since first-principles approaches would be ineffective, 
we elected to use an inverse data-driven method. Specifically, MOFAT utilizes \jwst spectra to drive full radiation transport simulations to converge on a set of parameters associated with the molecular forming region that best reproduces the observed molecular profiles.  
% By recreating \jwst spectra, MOFAT will be able to describe the scales of the 3D structures of the molecular region, effectively acting as a constraint to the underlying physical mechanisms that drive molecular formation and ultimately dust formation. 

\begin{table*}[]
\centering
    \caption{Parameter values from the best fit CO models of \citet{mera_jwst_2025} for the two epochs observed. Models P and O indicate the best-fitting models for prolate and oblate clumps, respectively. The temperature structures are shown in Figure \ref{fig:temp}. The \textit{Clumps} column indicates whether the model takes into account clumping effects. The \textit{Mass} column shows the total amount of CO mass. \textit{T$_1$} shows the temperature at the inner edge (\textit{v$_1$}) of the CO shell, which has a width of \textit{$\Delta$v}. \textit{n} is the slope of the CO density distribution. $r_R$ is the relative size of the clumps with respect to the radius. \textit{f$_c$} is the density enhancement factor within the clump with respect to its environment. $\epsilon$ is the flattening parameter of the spheroid shape of the clump (when $\epsilon =1$ it is a sphere, $< 1$ an oblate, and $ > 1$ a prolate). 
    }
\begin{tabular}{c|c|c|c|c|c|c|c|c|c|c}
\hline
CO Model & Time [d] &Clumps&Mass [M$_\odot$] & T$_1$ [K] & v$_1$ [$km~s^{-1}$] & $\Delta$v [$km~s^{-1}$]& n    & $r_R$     & f$_c$   & $\epsilon$ \\ \hline 
P  &285&yes   & 8.72e-3       & 2469      & 1240 & 3280 & 6.04 & 1.80e-1 & 1.16 & 70      \\ 
  &385&yes   & 1.31e-3       & 1951      & 1050 & 3700 & 6.64 & 1.38e-1 & 1.83 & 14      \\ 
  \hline 
O  &285&yes   & 8.88e-3       & 2423      & 1200 & 3210 & 6.86 & 1.72e-1 &  1.19 & 0.08      \\ 
  &385&yes   & 1.62e-3       & 1828      & 1130 & 3500 & 5.14 & 1.58e-1 & 1.89 & 0.46      \\
\hline
\end{tabular}

    \label{tab:model_values}
\end{table*}

In general, MOFAT assumes a spherically symmetric large-scale
geometry and uses a stationary solution of the radiation transport equations that can include the effects of multidimensional clump-like structures (see Appendix \ref{sec:global_rad}). To solve the radiation transport equations, we use formal integration in the observer's frame and assume a homologously expanding envelope with the molecular density distribution described by a power law ($\rho \sim r^{-n}$). In this solution, we use the so-called impact parameter ray system (see Figure \ref{fig:mofat} and \citealt{hummer_radiative_1971}) and rest-frame opacities corrected into the co-moving frame through the Sobolev approximation \citep{sobolev_moving_1960}. We emphasize that our implementation of this system uses modules developed within HYDRA \citep{hoeflich_core_2023, jwst_techniques_hoeflich_2024, numerical_hoeflich_2025}.
To calculate the rest-frame opacities, we use a given abundance ratio of C/O/Si and then utilize the molecular networks of HYDRA when given a temperature structure \citep{hoflich_infrared_2002,hoflich_multi-dimensional_2009,hoeflich_explosion_2017,hoeflich_measuring_2021}. 

The temperature structure, $T(v)$, in the molecular-rich region with expansion velocities $v~ \epsilon~ [v_1,v_1 +\Delta v]$ is obtained under the assumption of radiative equilibrium being established by the cooling (and heating) of molecules. In principle, $T(v)$ can be found by the formal solution involving the energy density in the co-moving frame. This method would quickly become unstable in MOFAT because it would require partial derivatives for each of the seven parameters (four in the non-clumping case). Instead, the optimal\footnote{Optimal means that $T(v)$ is found such that $\int \chi^2(\lambda) d\lambda$ between the observed and synthetic profiles is minimal.} temperature structure is calculated using a data-driven approach for a given set of parameters. 

The temperature at the inner boundary of the molecular region, $T(v_1)$, can be determined by the overtone bands. As we progress outward ($v_1<v<v_1 +\Delta v$), the temperature change is calculated based on the assumption that the total observed flux in the molecular bands must be produced in the molecular region. To do this, we utilize $T(v)=T(v)+ C ~\Delta T(v)$ to force agreement with the flux observed as we approach the outer boundary (see Appendix \ref{sec:temperature} for more details). For a given set of parameters, $C$ will approach unity with a dispersion comparable to the noise in the observations when the most optimal shape and strength of $\Delta T(v)$ is found\footnote{Some solutions of $\Delta T$ may be nonphysical.}. 

To describe the scales of instabilities, we include spheroidal clumps randomly situated throughout the observed molecular forming region. Their radiation effects are included in the general radiation transport equations through probabilistic encounters and a quasi-plan-parallel assumption (see Appendix \ref{sec:mol_region} and Figure \ref{fig:mofat}). These clumps are described by their shape or flattening parameter ($\epsilon$), their relative size to their radial dimension ($r_R$), and their density enhancement factor ($f_c$). Together, these parameters effectively control the surface to volume ratio and optical depth of the clumps, allowing for a `picket-fence' paradigm for the optically thick fundamental bands while retaining the `semi-'optically thin nature of the overtone bands. This picket-fence ability gives greater mobility to reproducing better flux ratios between the first and second vibrational modes, though the differences in spectra between oblate or prolate clumps may not be inherently distinguishable without sufficient time coverage \citep{mera_jwst_2025}.

\begin{table*}[]
\centering
    \caption{Parameter values from Figure \ref{fig:sio_ggi} optimized for the SiO profiles for the two epochs observed. The \textit{CO Model} column tells us the model for the initial temperature profile (see Figure \ref{fig:temp}). The rest of columns are the same as Table \ref{tab:model_values}. The SiO fundamental band is semi-optically thick during this time and shows shallow convergences, meaning that the clumping parameter results are inconclusive. For SiO model labeling we use the model of the initial temperature structure first and then the type of model used for the SiO fit given after. For example, if we use the temperature structure from Model P and then use oblate clumps (O) for the SiO fits then the label on the plots and in the text would be Model PO).
    }
\begin{tabular}{c|c|c|c|c|c|c|c|c|c|c|c}
\hline
CO Model & SiO Model & Time [d] &Clumps&Mass [M$_\odot$] & T$_1$ [K] & v$_1$ [$km~s^{-1}$] & $\Delta$v & n    & $r_R$     & f$_c$   & $\epsilon$ \\ \hline 
P & P  &285&yes   & 5.87e-3       & 2346      & 1430 & 1980& 1.97 & 4.50e-2 & 2.38 & 47      \\ 
  &&385&yes   & 2.32e-4 & 1925 & 1100 & 1970 & 7.30 & 1.00e-1 & 2.00 & 50      \\ 
\cline{2-12}
&O  &285&yes   & 3.92e-3       & 2423      & 1300 & 1700 & 1.83 & 1.00e-1 &  5.40 & 0.06      \\ 
  &&385&yes   & 4.95e-4       & 1925      & 1100 & 1600 & 2.37 & 9.00e-2 & 5.17 & 0.09      \\

\cline{2-12}
&S   &285&no  & 1.24e-3       & 2328      & 1700 & 500 & 9.88 & - & - & -      \\ 
&&385&no& 7.89e-5       & 2021      & 1030 & 715 & 10.17 & -       & -    & -       \\
\hline
\hline
O & P  &285&yes   & 2.62e-3       & 2314      & 1430 &1680 & 6.10 & 4.50e-2 & 7.00 & 100      \\ 
  &&385&yes   &   2.32e-4     & 2126      & 810 & 1450 & 1.10 & 2.08e-2 & 7.00 & 50      \\ 
\cline{2-12}
&O  &285&yes   & 2.13e-3       & 2507      & 1100 & 1670 & 5.16 &1.00e-1 &  1.50 & 0.07      \\ 
  &&385&yes   & 4.57e-4       & 1946      & 967 & 2450 & 1.91 & 5.50e-3 & 6.00 & 0.60      \\
\cline{2-12}
&S   &285&no  & 2.53e-3      & 2495     & 1700 & 575 & 10.30 & - & - & -      \\ 
&&385&no& 8.23e-5      & 1867     & 1050 & 750 & 10.30 & -       & -    & -       \\

 \hline
\end{tabular}

    \label{tab:siomodel}
\end{table*}

\begin{figure}
    \centering
    \includegraphics[width=0.9\linewidth]{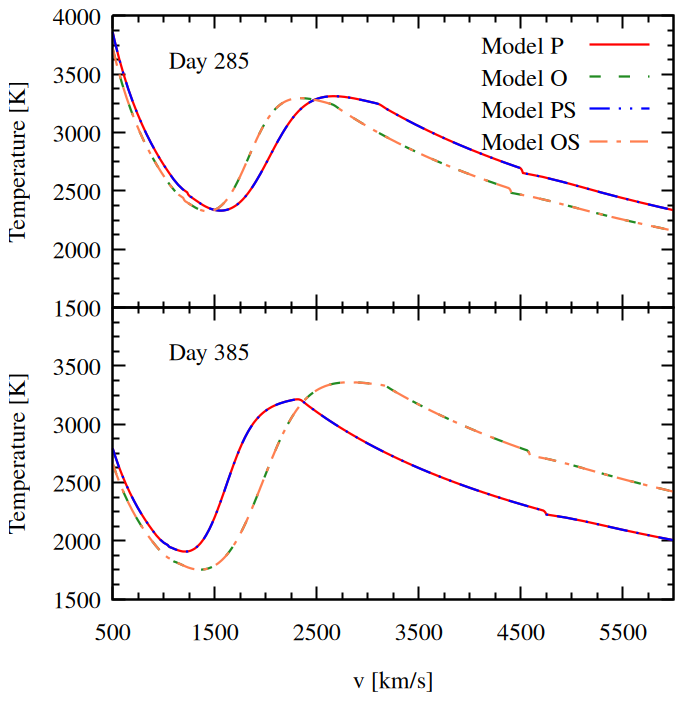}
    \caption{Temperature structures from select models from Tables \ref{tab:model_values} and \ref{tab:siomodel} at days 285 (top) and 385 (bottom). For all of our models, the SiO molecules do not significant cool the ejecta at either epoch.}
    \label{fig:temp}
\end{figure}

\subsection{Fitting Procedure} \label{sec:fitting_procedure}

The molecular forming region will have an outer edge that corresponds to the first instance of CO formation after the photosphere and recombination front have entered the C- and O-rich layers, and after the exposed region has sufficiently cooled to $\sim 5000$ K, the high-temperature limit for CO formation. As the inner edge of the molecular forming region follows the photosphere and recombination front, we expect CO molecules to form and further cool the surrounding envelope. When the CO molecules cool the envelope to temperatures around $\sim 2500$ K, SiO formation and cooling will begin in a Si- and O-rich region. This temperature cooling sequentialism and the assumption that the CO and SiO regions may have different multi-dimensionality effects (Section \ref{sec:motivation}) allow us to separate our fitting procedures into two steps --- 1) fit the CO features and find the overall temperature structure and then 2) use this temperature structure as starting solution when fitting the SiO features. 

To converge on an optimal parameter set, we utilize an iteration system of nested loops (see Appendix \ref{sec:iteration}), where we use a $\chi^2$ statistic for either the overtone or fundamental band to directly compare the fit between the synthetic and observed spectrum. We also employ an alternative $\chi_s^2$ statistic that measures the shape of the synthetic profile to the observations by scaling the absolute flux of our model to minimize the total flux difference with the observations. Hereafter, we use $\chi^2_*$ to indicate that either $\chi^2$ or $\chi^2_s$ can be used. The observed spectrum may contain overlapping line profiles, but we identify and include these line profiles in our continuum by taking the high-bandwidth
shape of the molecular bands to identify these low-bandwidth features. For any parameter variations, we automatically iterate the total molecular mass, NLTE strength, and temperature structure to optimize $\chi^2$ (see Appendix \ref{sec:temperature}). In general, total mass and NLTE strength are driven by the overtone. In the case where no distinguishable features are seen for the SiO overtone, we select a continuum lower than the observed overtone region to set an upper mass limit and then optimize the mass to the fundamental band. Based on 
$\chi^2$ and $\chi_s^2$, the sensitivity of specific features can be used to 
find an optimal parameter set for both the CO and SiO regions.

\subsubsection{Uniqueness of Solution}

The absolute flux ratio between the overtone and fundamental band, and the flux ratio between the vibrational modes in both allow for a unique solution. 
% With this fitting procedure, we directly overcome the initial difficulty that inhibits forward modeling from successful fits. That is, the ability to match flux ratios between the overtone and fundamental bands by correctly managing optical depths and cooling effects \citep{kotak_early-time_2005,mcleod_carbon_2024}. As a first-order step, this ratio completely rules out a large range of parameter space, effectively identified by MOFAT. The shape and quality of fit of parameter sets then act as the second-order driver for the uniqueness of solutions. 
For a given epoch, MOFAT can describe the molecular forming regions through seven converged parameters: the two radial boundaries ($v_{1}$ and $\Delta v$), three for the clumps (the size ratio of the clump to the radial dimension, $r_R$, the flattening parameter, $\epsilon$, and the density contrast with respect to the environment, $f_c$), the inner temperature edge\footnote{For SiO this parameter is synonymous with $v_1$ due to the starting $T(v)$ solution from the CO fit.} ($T_1$ defined at $v_{1}$), and the molecular density distribution ($\rho \sim r^{-n}$). CO and SiO regions will have two different optimized parameter sets, but will share $T(v)$. Together, these parameter sets can describe the structure of the molecular forming regions in SNe for a given spectrum. 
% For these seven parameters we utilize a fixed point iterative function to converge on the most optimal set of parameters via a $\chi^2$ statistic (see Section\ref{sec:iteration} and Figure \ref{fig:mofat_iteration}). The $\chi^2$ statistic is based on the fit between synthetic spectra and the actual observation for either the overtone or fundamental bands, where careful selection is paid to the more sensitive feature, but interchanging which feature to test is iterated for some parameters. 
With time series observations, we can follow the evolution of the molecular region through changes in the sets of converged parameters.

\begin{figure*}
    \centering
    \includegraphics[width=0.99\linewidth]{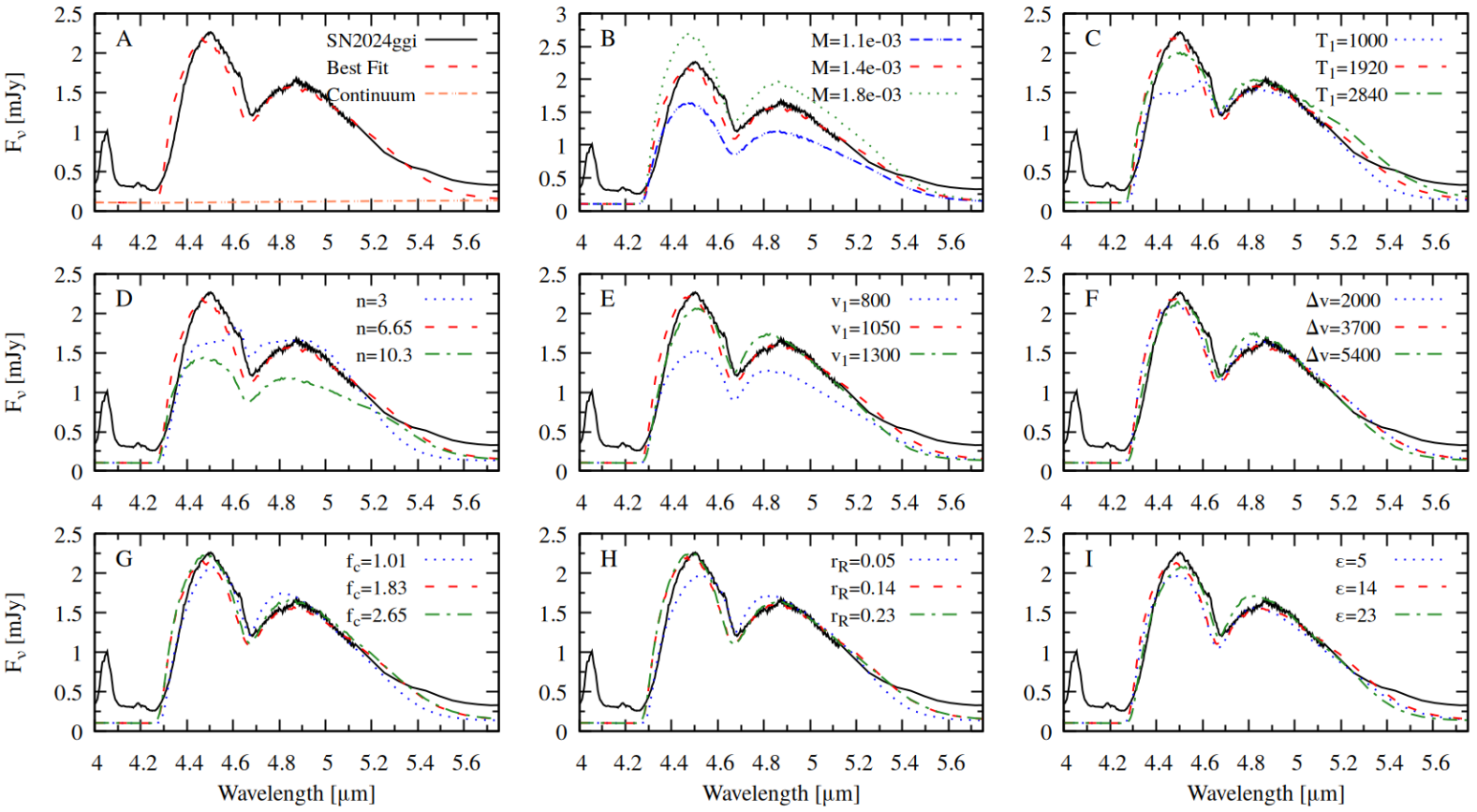}
    \caption{Demonstrating the sensitivity of each parameter on the CO fundamental band of Model P at day 385 (see Table \ref{tab:model_values}), where the temperature profile has been optimized for each. Panel A) shows the best-fit solution and continuum. Each of the following panels keeps the optimized parameters the same but shows the effect of the: B) total molecular mass ($M$), C) inner temperature ($T_1$), D) density structure ($n$), E) inner velocity edge ($v_1$), F) width of molecular region ($\Delta v$), G) clump density contrast ($f_c$), H) clump size ($r_R$), and I) clump shape ($\epsilon$). Figures \ref{fig:co_chi} and \ref{fig:co_chi_shape} show the $\chi^2_*$ convergences of the fits.}
    \label{fig:co_spectrum_fundamental}
\end{figure*}

\begin{figure*}
    \centering
    \includegraphics[width=0.99\linewidth]{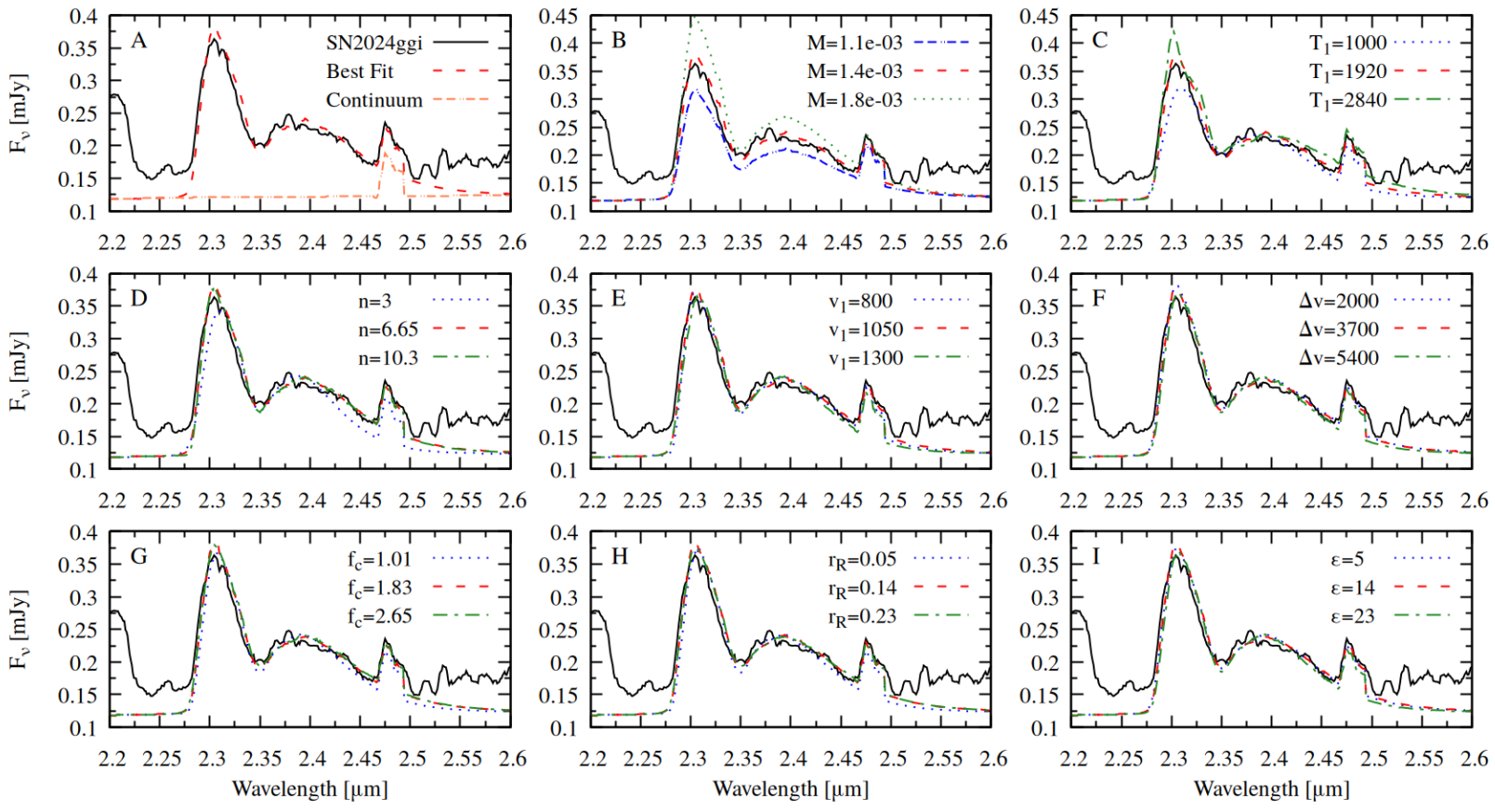}
    \caption{Same as Figure \ref{fig:co_spectrum_fundamental} but with the CO overtone band. Total mass (B) and temperature (C) influence the shape of the overtone more than the other parameters. In panel A, we show our continuum and have identified a low bandwidth Mg line profile at $2.475 ~\mu m$. Changes in the flux level of the continuum will only affect the total mass and temperature profile as they determine the absolute flux levels. }
    \label{fig:co_spectrum_overtone}
\end{figure*}

\section{Results: Application of MOFAT to \ggi} \label{sec:results}
In the following, we want to demonstrate the sensitivity of
the CO and SiO features to the free parameters and evaluate their solutions. We will use the NIRSpec and MIRI-LRS observations of \ggi\ first presented in \cite{mera_jwst_2025} and \cite{dessart_optical--infrared_2025}, respectively, at $+285$ and $+385$ days post-explosion. We use \ggi\ for its rapid classification \citep{hoogendam_ggi_2024} and well understood photospheric phase \citep{baron_jwst_2025} and because \cite{mera_jwst_2025} already showed MOFATs best-fit solutions for the CO features and found that clumping models (Models P and O, see Table \ref{tab:model_values} for the optimized parameter sets) are needed to explain the features at later epochs. Therefore, for the analysis of the sensitivity of the CO features, we use Model P at day 385 as a reference. We will then take the temperature structures of Models P and O at both epochs (Figure \ref{fig:temp}) and use them as a starting solution when we apply MOFAT to the SiO features of \ggi. Our aim for these new models is to analyze the sensitivity of fits and discuss their evolution over time to: (a) quantify total SiO mass, (b) assess the temperature and distribution of the corresponding stellar layers, and (c) investigate any small-scale 3D morphologies.

\subsection{Sensitivity of parameters on CO features and evaluation of reference model}\label{sec:paramter_study_co}

In Figures \ref{fig:co_spectrum_fundamental} and \ref{fig:co_spectrum_overtone}, we show the sensitivity of the parameters on the CO fundamental and overtone profiles, respectively. To quantitatively measure these variations against \ggi\ we show $\chi^2$ and $\chi_s^2$ convergence plots in Figures \ref{fig:co_chi} and \ref{fig:co_chi_shape}. Note that these figures illustrate variations around the best-fit solution, which was obtained by first applying the iteration scheme described in Appendix \ref{sec:iteration} (see also Figure \ref{fig:mofat_iteration}) over a broad parameter space grid. Potential alternate parameter solutions and correlations are discussed in Appendix \ref{sec:correlation}. The parameter ranges shown in these plots are the result of MOFATs initial wide search to constrain the range of possible parameters through the absolute flux ratio between the overtone and fundamental band. This overcomes the difficulty that forward-modeling simulations have had in producing high-fidelity fits to molecular features and allows for the study of the sensitivity of variations in parameters.

\begin{figure}
    \centering
    \includegraphics[width=0.99\linewidth]{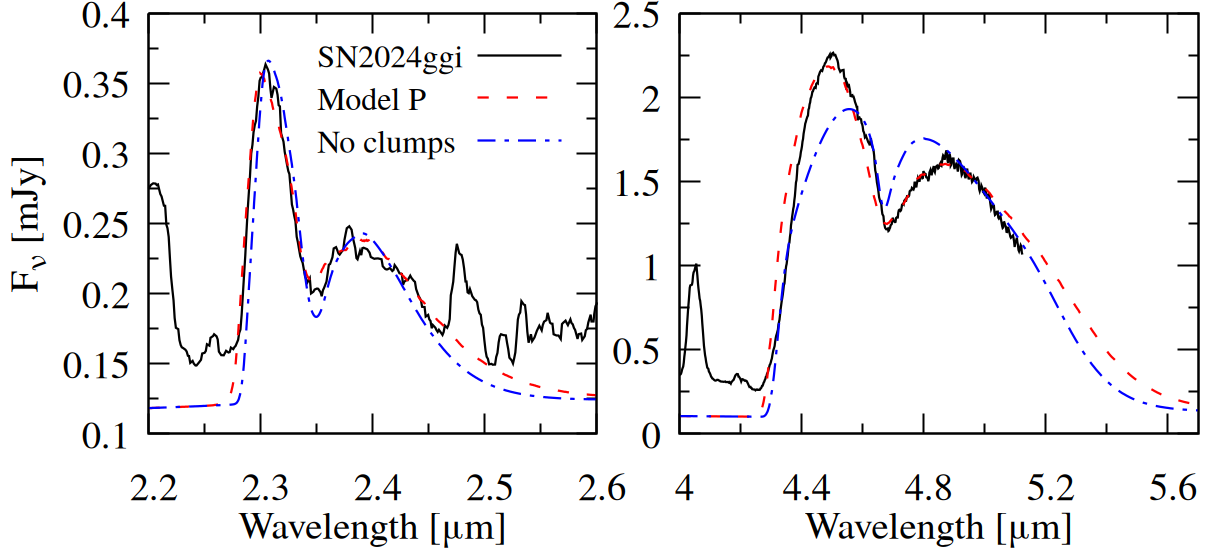}
    \caption{Spectrum of Model P with and without clumping.}
    \label{fig:model_p_no_clump}
\end{figure}

The interpretation of the variations depends sensitively on the overall temperature structure and on the optical depths of each band.
For our reference model, the optical depths are ray dependent; therefore, we show the optical depth for the first and second vibrational modes of both the overtone and fundamental features
in Figure \ref{fig:tau_co} for our reference model without clumps. To further illustrate the sensitivity of optical depth effects, we show the synthetic flux of our reference model with and without clumps in Figure \ref{fig:model_p_no_clump}. 
The overtone feature can be understood as a 
proxy for optically thin features, whereas the fundamental CO band represents a case of high optical depth and will experience more changes in the flux ratio between its first and second vibrational modes, as well as in the absolute flux due to variations in parameter sets with and without clumps.

\subsubsection{Total CO Mass}
In Panel B of Figures \ref{fig:co_spectrum_fundamental} and \ref{fig:co_spectrum_overtone}, we show the effect of total molecular mass when the temperature profile is kept constant. The overtone profile is more sensitive to the total mass because it is optically thin, so all CO mass will contribute. The fundamental band experiences a similar effect, but when there is less mass, the optical depth is reduced, allowing for better retention of the profile shape, although it deviates from the absolute flux of the observation. Among any parameter variations or temperature profile changes, the total mass is iterated to optimize the fit to the overtone. In some instances, parameter sets may cause the overtone band to become semi-optically thick, leading to higher total masses when compared to optically thin models, which may or may not have a higher $\chi^2$. If necessary, MOFAT will compensate for this by converging to more appropriate parameter regions.

\subsubsection{Temperature}
To first order, the emissivity between vibrational modes for the overtone bands reflects the strong temperature dependence 
of the opacity (see Panel C of Figure \ref{fig:co_spectrum_overtone} and
\citealt{hoflich_spectral_1988,liu_carbon_1992,rho_near-infrared_2021}). When the overtone begins to approach temperatures $\lesssim2100$ K, the first and second vibrational modes dominate. \ggi\ shows prominent first and second modes; therefore, we see a flattening of the $\chi^2$ and $\chi^2_s$ curves in Panel A of Figures \ref{fig:co_chi} and \ref{fig:co_chi_shape} for temperatures in this region. One might argue that the converged solution for the overtone sits around $1400$ K, but when the temperatures drop below $1700$ K, the strength of the vibrational modes of the fundamental band is no longer dominated by their statistical weights. As a consequence, the opacities of the first and second vibrational modes approach equality for this temperature region (see Panel C of Figure \ref{fig:co_spectrum_fundamental}). Therefore, at lower temperatures, the $\chi^2$ statistic for the fundamental band sharply increases for \ggi\ at $1600$ K (see Panel A of Figure \ref{fig:co_chi}). Comparing the $\chi^2$ and $\chi^2_s$ values for both the CO overtone and fundamental bands, the best fit solution for $T_1$ sits in the range between $\sim1800$ and $\sim2100$ K, with our reference model converging on a value close to the middle ($T_1=1951$ K).

\subsubsection{CO Density Distribution} 
\begin{figure*}
    \centering
    \includegraphics[width=0.99\linewidth]{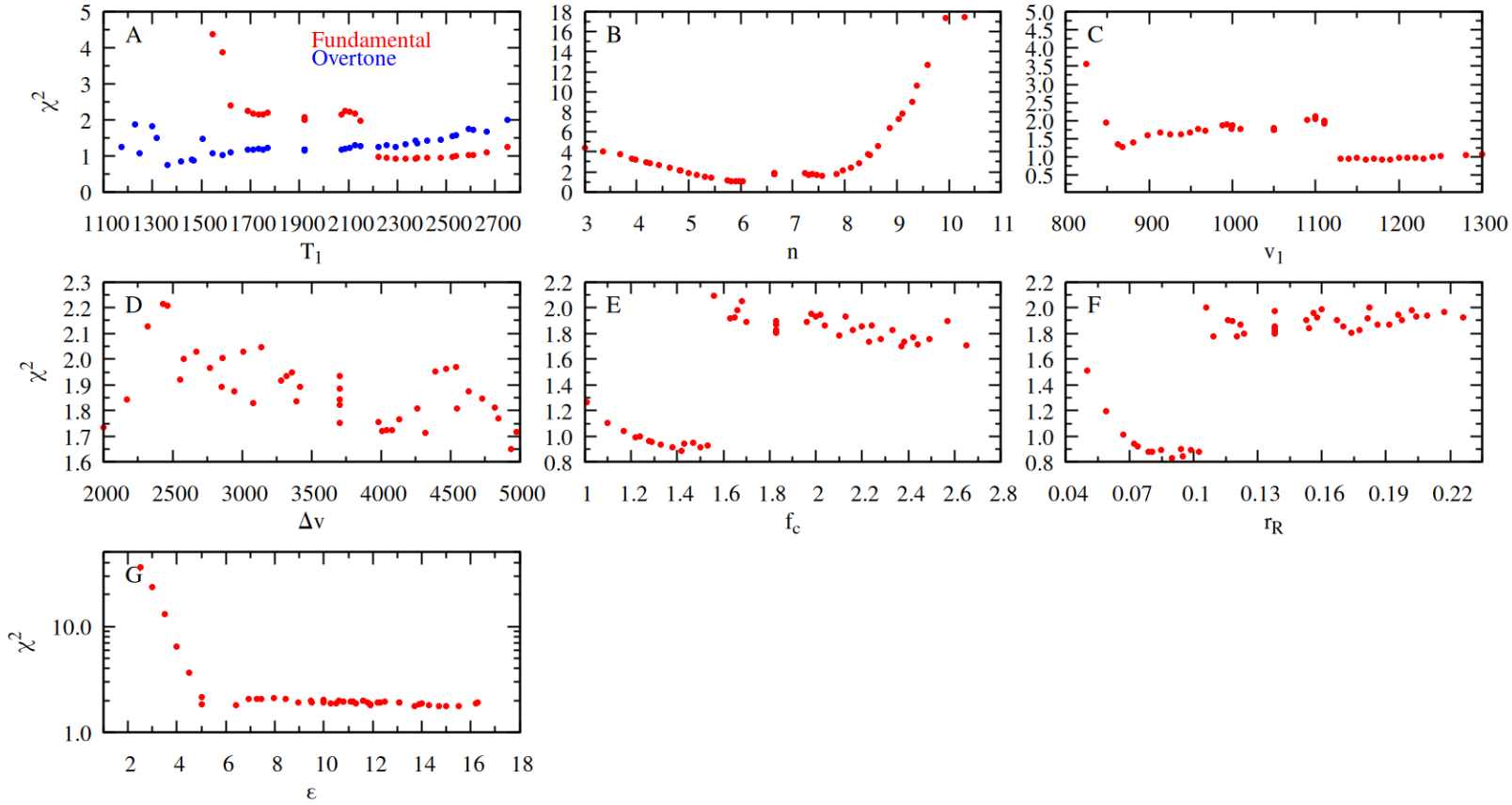}
    \caption{Demonstrating the $\chi ^2$ convergence for each of the seven parameters for Model P at day 385. For each panel, we keep the optimized parameters the same except for the parameter being varied. Panel A) shows temperature ($T_1$) convergence for both the CO fundamental and overtone bands. For panels B-G we only show $\chi^2$ of the fundamental band because the overtone does not show significant $\chi^2$ differences between models (see Figure \ref{fig:co_spectrum_overtone}). Panel B) shows density structure ($n$), C) inner velocity edge ($v_1$), D) width of the molecular region ($\Delta v$), E) clump density contrast ($f_c$), F) clump size ($r_R$), and G) clump shape ($\epsilon$). For discussion on the shapes and jumps, see Section\ref{sec:paramter_study_co}.}
    \label{fig:co_chi}
\end{figure*}

The density distribution determines the decoupling region for 
molecular areas that are partially optically thick and can directly change the flux ratio between the first and second vibrational modes of the fundamental band. The overtone is largely unaffected but may experience these effects in extreme cases. As the density slope approaches lower values, the radial separation between the decoupling regions for the first and second vibrational modes approaches zero, and the flux ratio becomes the same. This effect can clearly be seen for our fits of \ggi\ in Panel D of Figure \ref{fig:co_spectrum_fundamental}. Moreover, when the density slopes approach higher values, the decoupling regions are forced to lower velocities and are more clearly separated, allowing for more distinguished flux ratios between the vibrational modes. However, this may shorten the radial separation of the decoupling regions for the overtone and fundamental bands potentially triggering conflict when correcting the absolute flux through temperature corrections. For example, the higher density slope models for \ggi\ were unable to capture the absolute flux of the fundamental band because the temperature structures in the inner region are conserved due to the overtone (see Panel D of Figure \ref{fig:co_spectrum_fundamental}). From Figures \ref{fig:co_chi} and \ref{fig:co_chi_shape}, MOFAT converges close to the most optimal value of $n\approx6$. 

With time series observations, the change in density slope could mean one of two things. In the early stages of CO formation, the density slope is more closely related to the chemical abundance profile of the progenitor. However, at late times, after CO formation, the change in density slope demonstrates the effect that the external radiation field has on the destruction of the molecules \citep{mera_jwst_2025}. 

An important note to consider is that if the fundamental band is in a regime where it is not optically thick, then the convergence of the density structure would be uninformative, and it would be advisable to assume a density distribution from stellar models.

\subsubsection{Expansion Velocities}

For the CO overtone band, the expansion velocity results in small overall shifts or widening of the individual modes as the velocity increases (see Panels E and F of Figure \ref{fig:co_spectrum_overtone}). Although not present in \ggi, higher expansion velocities would smear out the individuality of the overtone vibrational modes at higher temperatures. 

In general, the fundamental band experiences similar shifts but will yield more nuanced results due to optical depth effects. 
As the velocities are reduced, the optical depth increases, reducing the overall absolute flux and flux ratio between the first and second vibrational modes (see Panel E of Figure \ref{fig:co_spectrum_fundamental} and the increase in $\chi^2$ and $\chi^2_s$ for $v_1\lesssim900$ $km~s^{-1}$). As expansion velocities increase, the optical depth decreases, which allows more flux from the central regions; however, the peak of the first vibrational mode is semi-optically thick, which still permits some blocking and a slight change in shape. In Figure \ref{fig:co_chi}, higher expansion velocities converge more for \ggi, but when cross-checked with the shape of the feature in Figure \ref{fig:co_chi_shape}, it is clear that the best converged value would fall somewhere between $900\lesssim v_1 \lesssim1050$ $km~s^{-1}$, with MOFAT converging on the high end.

\begin{figure}
    \centering
    \includegraphics[width=0.9\linewidth]{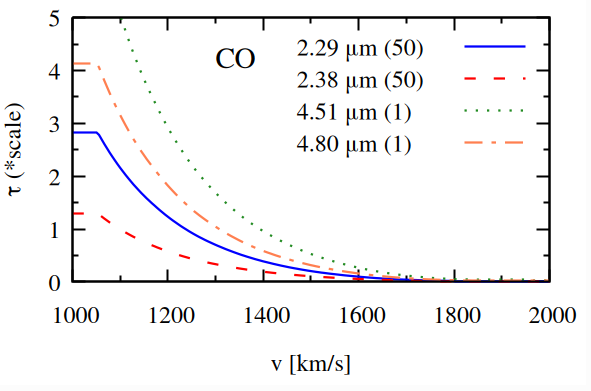}
    \caption{Optical depth of Model P at day 385 for the local peaks of the vibrational modes of the CO overtone (scaled) and fundamental bands if clumps are not included. If clumps are included then the optical depth is ray dependent. }
    \label{fig:tau_co}
\end{figure}

The width of the molecular region holds similar interpretations because, as the width shrinks, the optical depths in the fundamental bands increase, changing the overall profile shape. Although Figure \ref{fig:co_chi_shape} seems to converge somewhere between $2500\lesssim\Delta v \lesssim 4000$ $km~s^{-1}$ for \ggi, the high spread in variability results from the probabilistic encounters with the clumps, as well as from the idea that the optical depths of the first and second vibrational modes from Figure \ref{fig:tau_co} are well below $\tau<0.1$ at $\sim2000$ $km~s^{-1}$. This effect is also clearly seen in the flatness and spread of $\chi^2$ in Figure \ref{fig:co_chi}. For \ggi, we found that we require widths above $\Delta v\gtrsim2000$ $km~s^{-1}$ to conserve the shape and flux ratio of the two vibrational modes, but for widths larger than that, the interpretation becomes ill-defined. MOFAT finished with $\Delta v=3700$ $km~s^{-1}$, but a more appropriate value may be smaller.

\subsubsection{Clumps}
The three clump-related parameters ($f_c,~r_R,~\mathrm{and}~ \epsilon$) strongly affect the fundamental band while largely leaving the overtone band unaffected. These three parameters closely control the optical depth within the clump and the ability for the flux to escape through these picket-fence gaps. In broad terms, whenever these parameters increase the optical depth within the clump, the clump number density within the envelope decreases, where the amount of flux to escape from the central region will be determined by the overall clump shape. The implications are either --- 1) the increase in optical depth will reduce the flux ratio between the first and second vibrational modes, where the flux from the inner regions will increase as the relative clump surface area (or cross-section) becomes small, or 2) the decrease in optical depth will help retain the flux ratio between the vibrational modes, but the increase in clump number density may block flux if the relative clump surface area is high.

\begin{figure*}
    \centering
    \includegraphics[width=0.99\linewidth]{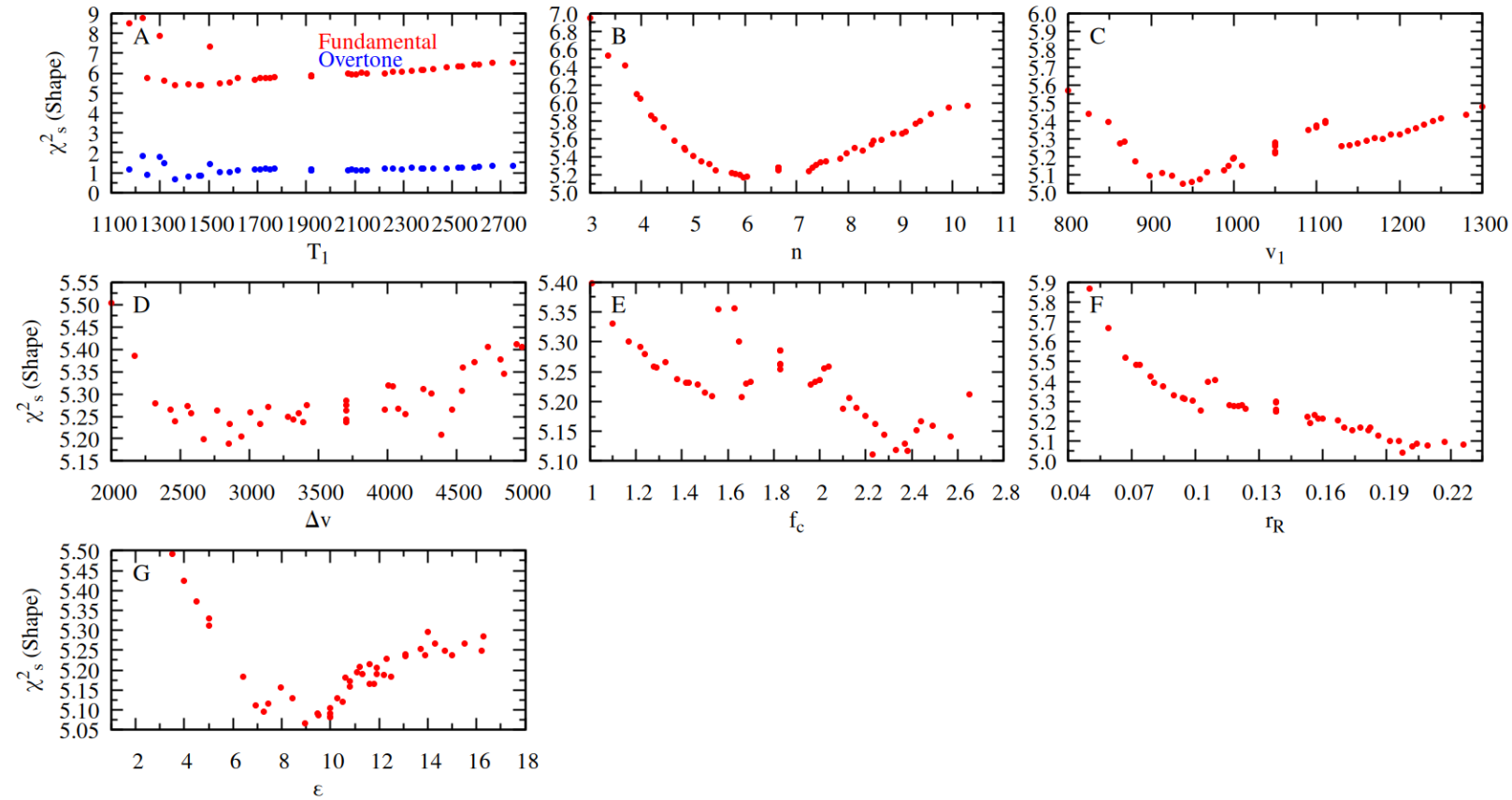}
    \caption{Same as Figure \ref{fig:co_chi} except that we scale the absolute flux of our model to minimize the total flux difference with the observation and then calculate $\chi^2_s$. This measures the relative shape of the model compared to the observations. }
    \label{fig:co_chi_shape}
\end{figure*}

For $r_R$, the smaller clumps allow more flux to escape from the inner region, producing more sharpened features (see Panel H of Figure \ref{fig:co_spectrum_fundamental}). Although these lower values seem to give more optimal $\chi^2$ fits in Panel F of Figure \ref{fig:co_chi}, the increase in optical depth within the clump blocks some of the flux from the first vibrational mode, making the shape become less optimal when compared to higher $r_R$ (see Panel F of Figure \ref{fig:co_chi_shape}). Moreover, the jump in $\chi^2$ at $r_R\approx0.1$ shows a sharp change in the optical depth of the clump. When we discuss the evaluation of the solutions for \ggi, we would expect the solution to converge somewhere close to this jump, as it is the region that best captures the ability of the clump to allow flux from the inner regions, while still being semi-optically thick to retain its spectral shape. MOFAT converged to $r_R$=0.14, a value well positioned when comparing the $\chi^2$ and $\chi^2_s$ convergences in Figures \ref{fig:co_chi} and \ref{fig:co_chi_shape}.

The clump density contrast parameter, $f_c$, in Panel G of Figure \ref{fig:co_spectrum_fundamental}, follows effects similar to $r_R$. For lower $f_c$, the optical depth within the clump decreases, but the increase in number density raises the probability of clump encounters. Because the feature is semi-optically thick, the flux from the first vibrational mode is partially blocked. As $f_c$ increases, the optical depth in the clump increases, but the decrease in number density allows for the increase of flux from the inner regions and the conservation of the flux ratio between the first and second vibrational modes. From Panel E of Figure \ref{fig:co_chi}, the change of optical depth effects occurs at $f_c\approx1.6$, but because of the shape conservation for increasing $f_c$ (Panel E of Figure \ref{fig:co_chi_shape}), an ideal solution would sit at a level slightly above this jump. This jump location may be specific for \ggi\ and can change for other SNe, but MOFAT was able to converge to a close value at $f_c=1.8$.

The shape, $\epsilon$, of the clump will not change the clump number density to a huge extent, therefore it is a direct measure of the volume to surface ratio of the clump. When the clumps are more sphere-like ($\epsilon=1$), the optical depths within the clumps is maximized which will reduce the absolute flux and the flux ratio between the two vibrational modes (see $\chi^2$ and $\chi^2_s$ for $\epsilon\lesssim6$ in Figures \ref{fig:co_chi} and \ref{fig:co_chi_shape}). As the shape becomes less spherical, the optical depth in the clump is decreased, and we see an increase in flux from the central region. However, as the clump shape becomes more irregular the clump surface area increases, which will allow for more clump interactions. This will block more flux from the first vibrational mode. These effects deliver a strong convergence of $\chi^2_s$ around $\epsilon\approx9$ in Panel G of Figure \ref{fig:co_chi_shape}. However, if we cross-check this solution with $\chi^2$ in Panel G of Figure \ref{fig:co_chi} we see that $\chi^2$ becomes relatively flat for $\epsilon>10$ until it reaches its minium at $\epsilon=15$ before increasing again. Compared together, we would expect a solution in a range between $10\lesssim\epsilon\lesssim13$. MOFAT finished with $\epsilon=14$ for \ggi, but with more iterations and more swapping between which $\chi_{(s)}^2$ convergence criteria to use, the best solution would most likely fall towards these lower values. Note, that this discussion holds similar value for oblate clumps ($\epsilon<1$), with potential differences caused by variations in surface to volume ratios.

\subsection{Analysis of SiO features of \ggi} \label{sec:sio_fits}

\begin{figure*}
    \centering
    \includegraphics[width=0.95\linewidth]{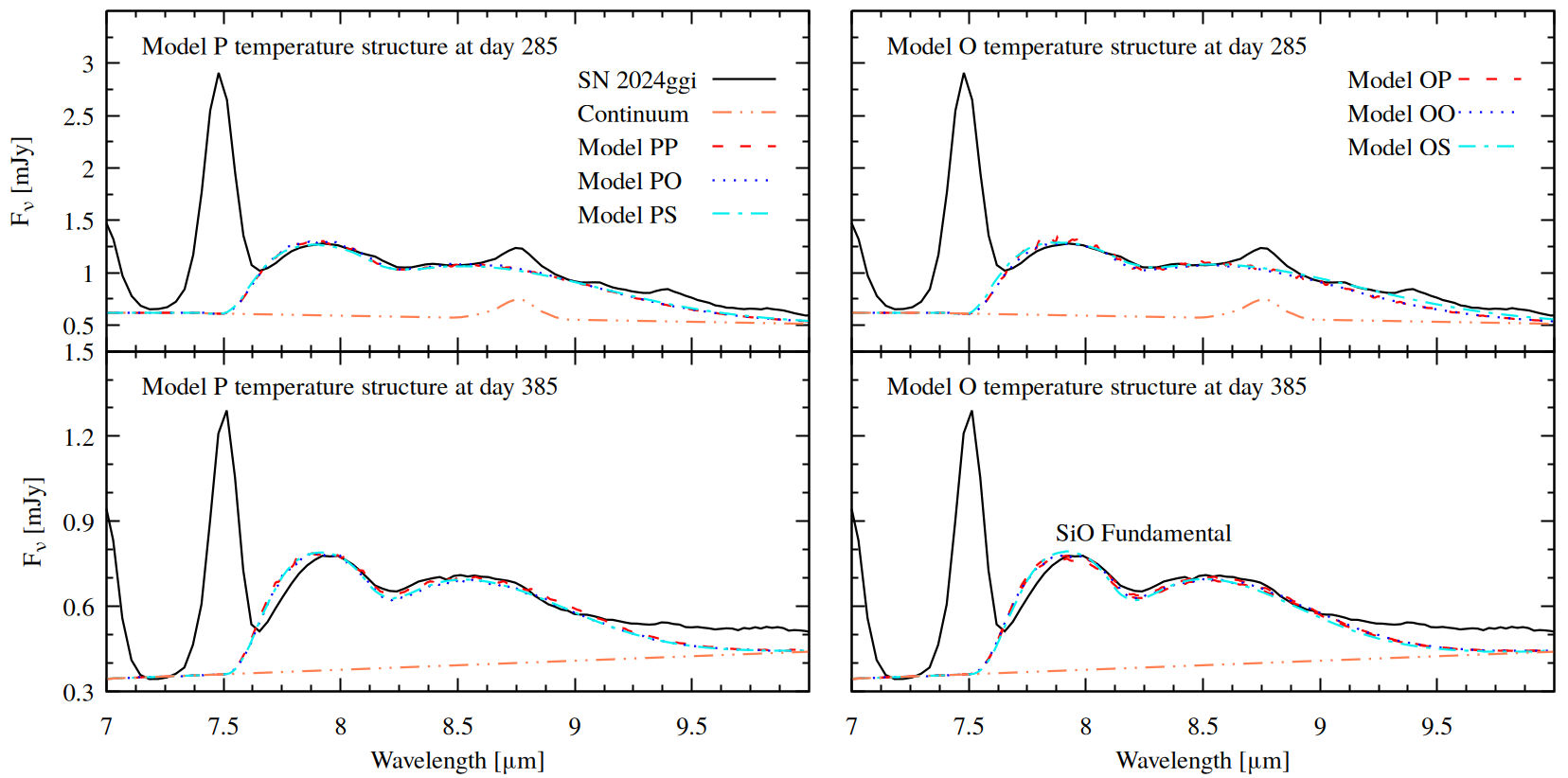}
    \caption{For given initial temperature structures of Models P (left) and O (right, Figure \ref{fig:temp}) we show our best SiO fundamental bands fits for prolate (P), oblate (O), and non-clumping (S) models. The continuum used for day 285 shows the identification of a low bandwidth H line profile at $8.75~ \mu m$ that overlaps our high-bandwidth molecular feature. Also included in the continuum but not plotted is the edge of the [Ni I] line at $7.5~ \mu m$. For parameter values, see Table \ref{tab:siomodel}. } 
    \label{fig:sio_ggi}
\end{figure*}

In Figure \ref{fig:sio_ggi}, we show our best fit models for the SiO features of \ggi\ at both epochs using the temperature profiles from Models P and O as starting solutions for our prolate (P), oblate (O), and non-clumping models (S). All converged model parameters are given in Table \ref{tab:siomodel}. Across all models, we find no evidence for SiO cooling (see Figure \ref{fig:temp}). Additionally, we find little evidence for the SiO overtone band in either epoch and only use that region to set an upper limit on the SiO mass, where the fundamental band is then used as a proxy for the total mass. In Appendix \ref{sec:sensitivity_sio}, for Model PP at day 385, we demonstrate the sensitivity of each parameter on the SiO features in Figure \ref{fig:sio_spectrum}, where we show the $\chi^2$ and $\chi^2_s$ convergence plots in Figures \ref{fig:sio_chi} and \ref{fig:sio_chi_shape}, respectively. 

\begin{figure}
    \centering
    \includegraphics[width=0.9\linewidth]{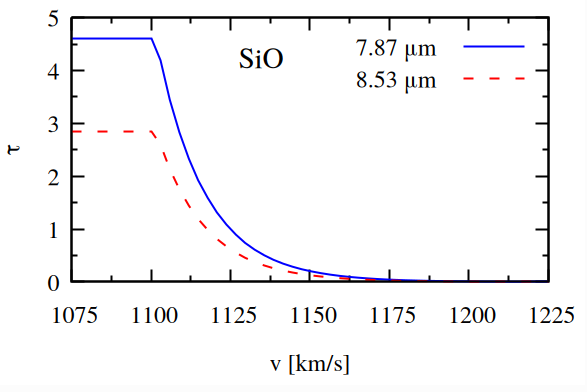}
    \caption{Optical depth of Model PP at day 385 for the local peaks of the vibrational modes of the SiO fundamental band if clumps are not included. }
    \label{fig:tau_sio}
\end{figure}

At these temperatures, SiO is expected to be optically thin, but in Figure \ref{fig:tau_sio} we show that the SiO fundamental band is semi-optically thick. This means that we may have the ability to probe clump-like structures as long as we are able to strongly converge on a density structure. In Panel A of Figure \ref{fig:sio_chi} we show that Model PP converges to $n\approx8$, but we caution that the convergence range is fairly flat for $6\lesssim n\lesssim11$, with $\chi^2_s$ showing a general downward trend as $n$ increases, a sign that the feature has low optical depth. The other parameters in Figure \ref{fig:sio_chi} tend to converge relatively well, but they also exhibit a rather wide and shallow range of potential solutions, with $\chi^2_s$ staying relatively flat in those parameter ranges. Therefore, we will adopt a conservative approach in the interpretation of the models, and will mainly discuss the general trends for non-clump-related parameters. 

At both epochs, all SiO models show relatively good agreement to the observations of \ggi. Several similarities are common between each model shown: the SiO-mass for all models decreases\footnote{If we assume a constant $n$ across both epochs for all models, we still see a decrease in overall mass}, with the amount of decrease ranging by factors between $\approx5$ and $30$; the location of the SiO rich region in velocity space decreases as time progresses; and the overall temperature of the SiO molecules decreases from $\approx2400$ K to $\approx1900$ K. We showed in Figure \ref{fig:temp} that SiO does not cool the ejecta during either epoch, which means that the CO molecules dominate the cooling during this time with temperatures still too high for dust formation. As a result, the formation of SiO is dependent on CO molecules to cool the envelope to temperatures below $\sim2500$ K, which can clearly be seen in narrow regions in Figure \ref{fig:temp}. The temperature minimum from the CO models coincides with the inner edge of the SiO region for both epochs, supporting the idea that Si is mixed into the C/O layers and requires CO cooling to initiate SiO formation. 

From \cite{mera_jwst_2025}, we learned that CO formation has most likely finished by day 285 and that geometrical dilution has exposed the CO molecules to hard radiation that led to CO destruction and re-heating of the outer molecular regions. As a consequence, we suspect two scenarios --- 1) SiO undergoes rapid formation timescales, or 2) CO cooling may have cooled larger regions before what we see in Figure \ref{fig:temp}, allowing for an adequate amount of SiO to form before day 285. However, because these outer regions are actively exposed to hard radiation SiO molecules are strongly dissociated as time progresses. Moreover, the progression of the inner SiO region to lower velocities invites the interpretation that we may still be in an active time-frame for SiO formation and within the Si- and O-rich layers.

We have demonstrated the sensitivity of the parameters on the CO molecular profiles and shown their ability to converge on an optimal parameter set when compared to the observations of \ggi. We then applied MOFAT to the SiO features of \ggi\ and found good agreement, with the general conclusion that we are actively forming but mostly destroying SiO, with inconclusive results regarding the presence of clump-like structures. For future observations, we need low-latency observations during the CO formation period to probe mixing lengths and cooling regions conducive for SiO formation, and observations at late times to study small-scale 3D morphologies in the SiO formation regions and to address whether SiO and CO cooling can trigger dust formation.

\section{Discussion}\label{sec:dis}

% In Section \ref{sec:model}, we presented our MOlecular Fitting Analysis Tool (MOFAT) that uses observations to drive time-independent radiation transport simulations to probe the 3D structures of the CO and SiO molecular forming regions in SNe. MOFAT was created to overcome the difficulties of complex 3D time-dependent SNe simulations by enabling many different envelope structures to be tested more efficiently (see Section \ref{sec:motivation}). Its parameter space can reproduce the flux ratios between the CO overtone and fundamental bands, as shown in Figures \ref{fig:co_spectrum_fundamental} and \ref{fig:co_spectrum_overtone}. 
In Section \ref{sec:model}, we presented our MOlecular Fitting Analysis Tool (MOFAT), which uses observations to drive time-independent radiation transport simulations to probe the 3D structures of CO and SiO molecular-forming regions in SNe. Unlike previous one-zone time-dependent simulations, which struggled to reproduce the flux ratios between the CO overtone and fundamental bands, MOFAT was developed to overcome these challenges by efficiently exploring a wide range of envelope structures and successfully captures these ratios and profile shapes across its parameter space (see Figures \ref{fig:co_spectrum_fundamental} and \ref{fig:co_spectrum_overtone}).
In \cite{mera_jwst_2025}, MOFAT was applied to \ggi\ at $+285$ and $+385$ days, and found that: CO formation has finished by day 285, CO is being actively destroyed, and that clumping structures are needed to explain the CO molecular features at late times. As supporting evidence, this paper demonstrated the sensitivity of each parameter and evaluated the quality of the solutions in Section \ref{sec:paramter_study_co}. We then applied MOFAT to the SiO features of \ggi\ in Section \ref{sec:sio_fits} and found that: SiO has formed prior to day 285 and requires CO cooling to form; it is not a dominant coolant, is being actively formed but mostly destroyed, and there are inconclusive results regarding the presence of clump-like structures.

MOFAT is able to recreate observations with high fidelity by using a combination of $\chi^2$ and $\chi^2_s$ statistics within an iteration system of nested-loops built to converge to the most optimal parameter set (see Appendix \ref{sec:iteration}). As the first step, MOFAT identifies any overlapping line profiles and includes them in the continuum, allowing for an extended wavelength range for our $\chi^2_*$ calculations. The convergence of a parameter set is contingent on an initial good guess because the parameters in the inner loops are extremely sensitive to the parameters in the outer loops (e.g., if the clump parameters encourage optically thick features, then MOFAT may drive the outer loop parameters to unlikely regions in order to recover flux ratios between the vibrational modes). If the parameters start in a regime where the clumps are partially optically thick, then MOFAT will be able to converge towards a solution, making it a powerful diagnostic tool.    

One of MOFATs primary diagnostic capabilities is its ability to use molecular features at high optical depths to probe and characterize small-scale 3D structures, providing a link to the instabilities from evolutionary or explosion physics (see Section \ref{sec:motivation}). MOFAT achieves this by using optically thick fundamental bands to test clump parameters until the total absolute flux and the flux ratios between vibrational modes optimally reproduce observations (see Section \ref{sec:paramter_study_co}). These clump parameters then describe the general instability scales in the molecular forming regions of an SN envelope,. However, distinguishing between prolate or oblate structures requires low-latency extended time-frame observations, which can change the interpretation of the instability type (i.e., RT instabilities during the explosion or even small-scale mixing between the Si- and C-rich layers by shear flows). As a corollary, when features are semi-optically thick or optically thin, the lack of evidence for small-scale structures does not imply their absence (see Section \ref{sec:sio_fits} and \citealp{mera_jwst_2025}). This emphasizes the need for good coverage of early time observations to distinguish the origin of clumps and late time observations to solidify their presence and evolutionary pathways.

Optically thin bands provide a sensitive diagnostic tool when determining the total mass of molecular species and the temperature of the inner region. The separation and flux height ratio of the vibrational modes of the overtone can be used as an indicator for the temperature, where the absolute flux is a direct measure of the total overall mass (see Section \ref{sec:paramter_study_co} and \citealp{rho_near-infrared_2021}). In the early stages of molecular formation, the fundamental band may be optically thin and can act as the main determiner for these attributes (see Section \ref{sec:sio_fits}). As a secondary diagnostic ability, the overall smearing between vibrational modes and the width of features will act as a measure for the expansion velocities, giving boundaries to the regions of molecular formation (see Section \ref{sec:paramter_study_co}).  

With time-series observations, the boundaries continuously set by CO and SiO regions provide the ability to use tomography as a tool for the reconstruction of the Si/O/C layers of the SN envelope, providing a direct link to the progenitor. The initial detection of CO would indicate the outer boundary of the C- and O-rich layers. The CO region will continuously extend towards lower velocities, where eventually the formation and cooling of CO will trigger SiO formation, thus revealing the outer boundary of the Si- and O-rich region. These regions can then be followed until their features are no longer visible, but their cooling effects may be seen in later observations with a rise in dust formation (see Section \ref{sec:intro}). This mapping of the chemical layers can give insight into additional mixing lengths, potentially separate from the length scales of the clumps, with origins from either evolutionary or explosive physics. The separation of these two length scales further emphasizes the need for low-latency extended time-frame observations that cover early CO formation all the way to dust formation.

\section{Conclusion}\label{sec:con}
We have introduced and explored in detail the ability of MOFAT to converge on optimal parameter sets to recreate CO and SiO profiles with high fidelity. We showed that these parameter sets can be used to decipher abundance and temperature structures and small-scale 3D morphologies of SNe through the use of optically thick and thin features (see Section \ref{sec:paramter_study_co}), providing a direct link to the progenitor and instabilities caused by evolution or explosion physics. 

MOFAT was applied to the CO features of \ggi\ in \cite{mera_jwst_2025} and to the SiO features in Section \ref{sec:sio_fits} at $+285$ and $+385$ days post-explosion. The results of these studies found complicated evolutionary paths for the clumping of the CO molecules, the destruction of both types of molecules (SiO is potentially still actively forming), and the reliance of CO cooling to trigger the formation of SiO during these epochs. The limited temporal coverage at early formation times, combined with the weak convergence of the SiO features, makes the application of tomography difficult. Nevertheless, the results and first applications of MOFAT demonstrate that it has the powerful ability to probe the 3D structures of SNe with sufficient time coverage. 

We caution, however, that there are significant limitations to MOFAT. The models discussed here assume large-scale spherical symmetry by assuming uniformly distributed clumps with the same relative sizes and shapes. From full-time-dependent radiation-hydrodynamic simulations and observations, we expect to see a large range of instability types with spectrum variations caused by the viewing angle (see Section \ref{sec:motivation} and \citealp{khokhlov_jet-induced_1999, couch_aspherical_2009,hoeflich_measuring_2021,hoeflich_core_2023,temim_dissecting_2024,vartanyan_3d_2025}). MOFAT can only measure the average of these scales or the most dominant one, which limits the interpretation of the instability type and any potential separation of clump shape. Ideally, the surface to volume ratio difference between prolate and oblate clumps will give measurable differences between the flux ratio of the first and second vibrational modes. Studies are currently underway to study these differences, but to accurately separate their interpretations, MOFAT would require high spectral coverage of the entirety of the CO (or SiO) formation period. Full 3D time-dependent simulations may be able to constrain clump types, but this would require a large grid of models and many well-sampled SNe. Lastly, MOFAT is only as accurate as the observations it uses. \ggi\ was a relatively close CC~SN with a strong S/N ratio, but if a SN is further away then the S/N ratio reduces the reliability of the converged parameter sets and any indication for clump-like structures because of the sensitivity of the solutions. 

Overall, we have shown the reliability of MOFAT when explaining the molecular forming regions of SNe with its inaugural application to the CO and SiO features of \ggi. Clump-like structures are mostly necessary to explain molecular features, but confidence in their presence require optically thick molecular bands. Looking ahead, extended time coverage observations of an SN would allow MOFAT to use tomography to recreate the envelope structure including chemical stratification, instability scales, and temperature structures. These structural maps can then be used to constrain the role molecules play in the formation and abundance of dust from SNe, helping resolve their influence on the dust budget of the Universe.  

\section{Acknowledgments}\label{sec:ackowl}
T.M. acknowledges the support of friends, family, and fellow collaborators throughout the entirety of his PhD, during which this work was conducted.

T.M., P.H., and E.F. acknowledge the support by the National Science Foundation NSF awards AST-1715133 and  AST- AST-230639 for enabling the development of the methods and computational codes.

T.M., K.M., E.B., C.A., M.S., and  P.H. acknowledge support from NASA grants JWST-GO-02114,
JWST-GO-02122, JWST-GO-04522, JWST-GO-04217, JWST-GO-04436,
JWST-GO-03726, JWST-GO-05057, JWST-GO-05290, JWST-GO-06023,
JWST-GO-06677, JWST-GO-06213, JWST-GO-06583, JWST-GO-0923. Support for
programs \#2114, \#2122, \#3726, \#4217, \#4436, \#4522,  \#5057,
\#6023, \#6213, \#6583, and \#6677
were provided by NASA through a grant from the Space Telescope Science
Institute, which is operated by the Association of Universities for Research in
Astronomy, Inc., under NASA contract NAS 5-03127.

W.B.H. acknowledges support from the National Science Foundation Graduate Research Fellowship Program under Grant No. 2236415. 

%%%%%%%%%%%%%%%%%%%%%%%%%%%%%%%%%%%%%%%%%%%%%%%%%%
\section{Data Availability}\label{sec:data_avail}
The observations of \ggi\ can be accessed via: \dataset[doi:10.17909/3v0v-0a76]{https://doi.org/10.17909/3v0v-0a76}. MOFAT will be available at \href{https://tycobrahe.github.io/}{tycobrahe.github.io} upon the completion of the thesis and will contain all models shown in this manuscript.

\bibliographystyle{aasjournalv7}
\bibliography{zotero,refs} 

\appendix
\section{Radiation transport with MOFAT:}\label{sec:global_rad}
% MOFAT assumes spherically symmetric large-scale
% geometry and uses a stationary solution of the radiation transport equations (radiation transport equation) that can include the effects of multidimensional clump-like structures.

% \subsection{Formal solution of the radiation transport for global spherical geometry}\label{sec:global_rad}
For the radiation transport equations, we use standard notation as in 
\cite{hummer_radiative_1971} and \cite{hubeny_theory_2014} and assume that our SN envelope is expanding homologously and isotropically. As stated above, we solve the time-independent spherically-symmetric radiation transport equations in the observer's frame and correct our opacities into the observer's frame. In principle, we could include scattering terms by solving the radiation transport equation in the co-moving frame in non-relativistic atmospheres \citep{mihalas_solution_1975}, however, we assume that our sources are pure blackbodies. 
% In most cases the radiation transport equation are given in the standard $(r,\mu)$ coordinate system
% The system contains a module to solve the radiation transport equation in the comoving frame in non-relativistic atmospheres \citep{mihalas_solution_1975}, which allows us to include the contribution from scattering terms --- these scattering terms may be treated as negligible because molecule and dust features are assumed to be in local-thermodynamical equilibrium (LTE).
% For the emergent flux, we use a formal integration in the observers frame. 
% The time-independent radiation transfer equation in a spherically symmetric atmosphere is:
% \begin{equation}
%     \left[ \mu\frac{\partial}{\partial r} + r^{-1}(1-\mu^2)\frac{\partial}{\partial \mu} \right]I(r,\mu,\nu) = \eta(r,\nu) - \chi(r,\nu)I(r,\mu,\nu),
% \end{equation}
% where $\eta(r,\nu)$ and $\chi(r,\nu)$ are the total emissivity and opacity, respectively. In the observers frame they become $\eta(r,\nu,\vec{v})$ and $\chi(r,\nu,\vec{v})$ --- obtained by the Lorentzian transformation based on the comoving frame values, which are assumed to be isotropic.
%  $\mu$ is the cosine of the angle between a radius vector and the direction of radiation propagation at r. If we transform from 
 Transforming from the standard $(r,\mu)$ coordinate system to the $(p,z)$ representation, i.e. the so-called impact parameter $p=r\mathrm{sin}(\theta)$, and the distance $z=\sqrt{r^2 - p^2}$ along the ray system (see Figure \ref{fig:mofat}), the radiation transport equation becomes:
\begin{equation}\label{eq:I(p,z)}
    \frac{1}{\kappa(p,z,\nu)}\frac{\partial I^\pm (p,z,\nu)}{\partial z} = I^\pm(p,z,\nu) - S(z,\nu),
\end{equation}
where 
$S(z,\nu)$ is the source function and $\kappa(p,z,\nu)$ are the co-moving frame opacities. 
% $S(r,\nu,\vec{v})=\kappa(r,\nu,\vec{v}) B(r,\nu)/(\kappa(r,\nu,\vec{v}) + \kappa_{Thompson})$. 
% $S(r,\nu,\vec{v})=\eta(r,\nu,\vec{v})/\chi(r,\nu,\vec{v})$ and have used the relation $\partial \tau = -\chi(r,\nu,\vec{v})\partial z$.
% The source function can be approximated by the contribution from thermal absorption and emission plus the contribution from Thomson scattering by free electrons:
% \begin{equation}
%     S(r,\nu,\vec{v})=\frac{\kappa(r,\nu,\vec{v}) B(r,\nu) + \sigma_{th}J(r,\nu)}{\kappa(r,\nu,\vec{v}) + \sigma_{th}} = \Theta(r,\nu,\vec{v})  B(r,\nu) + (1-\Theta(r,\nu))J(r,\nu,\vec{v}),
% \end{equation}
% where $B(r,\nu)$ is the blackbody function, $\kappa(r,\nu,\vec{v})$ are the observers frame opacities, and $\sigma_{th}$ is the opacity due to Thompson scattering. The solution to Eq. \ref{eq:I(p,z)} is then
% \begin{equation}
%     I(p,\tau_1,\nu) = I(p,\tau_2,\nu)e^{-(\tau_2-\tau_1)} + S(r,\nu,\vec{v})\int_{\tau_1}^{\tau_2}e^{-(t-\tau_1)}dt.
% \end{equation}
If we introduce a flux-like variable $\mathscr{V}(p,z,\nu)=\frac{1}{2}(I^+(p,z,\nu)-I^-(p,z,\nu))$, the formal flux integral is
\begin{equation}\label{eq:H}
    H(r,\nu)=\frac{1}{r^2}\int_0^rdpp\mathscr{V}(p,z,\nu).
\end{equation}
To solve this equation we use the radiation transport modules of HYDRA \citep{hoeflich_core_2023,jwst_techniques_hoeflich_2024,numerical_hoeflich_2025}, which have been well tested against many types of transients. 
\subsection{Treatment of molecular region in MOFAT} \label{sec:mol_region}
MOFAT can treat the molecular region as a diffuse homogeneous layer, or it can treat it as a region containing molecular clumps.
% Based on detailed, spherical but time-dependent simulations, CO/SiO is formed right below the recombination front in the C/O/Si core via e.g. $CO^+\rightarrow CO$ and close to the photosphere because of the quadratic dependence on the formation rate due to the density. 
% In both cases the outer edge of the CO/SiO-rich region corresponds to the earliest time of CO/SiO-formation and the inner edge follows the recombination front until the photosphere enters the Si- and O-poor layers. 
If clumps are absent, Eq. \ref{eq:H} is treated as is. If clumping is included, then the radiation transport is considered as the radiation transport between the clumps, meaning the absorption/emission effects along a ray must include the multidimensionality of the clump. For the radiation transport along a ray, the impact of the clumps are characterized by the probability to hit a clump for both emission and absorption processes. For the emission process, we assume a quasi-plan-parallel approximation to account for self-absorption effects within the clump. Both processes use the clumps effective optical depth.

\subsubsection{Clump parameters}
We assume our clumps are spheroids with a major axis, $r_c$, and a minor axis, $r_a$, which is proportional to the major axis and has a flattening parameter $\epsilon$ (it is oblate when $\epsilon<1$, a sphere when $\epsilon=1$, and prolate when $\epsilon>1$). We define $r_R$ as the relative size of $r_c$ to the radial dimension. The equations for the semi-axes of the clumps are:
\begin{equation}\label{eq:semi_axes}
    \begin{aligned}
    r_c(r) &= r_R~r \\
    r_a(r) &= \left\{ \begin{aligned} 
    &\epsilon r_c(r) ~~~~~~~~~~~~~~~~~~~~ \mathrm{for} ~~~ 0 < \epsilon \leq 1 \\
    &\frac{r_c(r)}{\epsilon} ~~~~~~~~~~~~~~~~~~~~~ \mathrm{for} ~~~ \epsilon >1 .
    \end{aligned}\right.
    \end{aligned}
\end{equation}

The volume of the clump at a given radius will then be:
\begin{equation}
\begin{aligned}
    V_c(r,\epsilon) &= \left\{ \begin{aligned} 
    &\frac{4\pi}{3}\epsilon r_c(r)^3 ~~~~~~~~~~~~~~~~~~~~ \mathrm{for} ~~~ 0 < \epsilon \leq 1 \\
    &\frac{4\pi}{3}\frac{r_c(r)^3}{\epsilon^{2}} ~~~~~~~~~~~~~~~~~~~~~ \mathrm{for} ~~~ \epsilon >1 
    \end{aligned}\right.
    \end{aligned}
\end{equation}

The surface area of the spheroid is proportional to $\mathrm{arctanh}(\epsilon)$; however, this is computationally expensive, so we take the approximate form. This is known as Thompson's formula and has a relative error $< 1.42\%$ \citep{michon_final_2022}:

\begin{equation}\label{eq:sa_sphrd}
    \begin{aligned}
            S_c(r,\epsilon) & \approx \left\{ \begin{aligned}
            &2\pi (r_c/ \epsilon)^2(1 + 2\epsilon^p)^{1/p}~~~~~~~~~~~~~~~~~~~~\mathrm{for}~~~ \epsilon > 1\\
             &2\pi r_c^2(1 + 2\epsilon^p)^{1/p} ~~~~~~~~~~~~~~~~~~~~~~~~~~~~ \mathrm{for}~~~ \epsilon \le 1
                        \end{aligned} \right.
    \end{aligned}
\end{equation}
where $p=\mathrm{ln}(3)/\mathrm{ln}(2)$. 

% \begin{figure}
%     \centering
%     \includegraphics[width=0.8\linewidth]{fig10.png}
%     \caption{Left: The coordinate systems of $(r,\theta)$ and $(p,z)$. $r_{photo}$ is the photosphere, $r_1$ and $r_1+\Delta r$ are the inner and outer radius of the molecular shell, respectively, and $r_j$ is some arbitrary radius. These locations coincide with the equivalent velocities in the right figure. Right: A cartoon graphic of how MOFAT works. The molecular clumps have semi-axes of $r_c$ and $r_a$ with a density enhancement factor of $f_c$. When we integrate along $p_j$, the ray will have a certain probability ($P$, yellow star) of hitting a clump at $v_j$. The effects of the clump are then included, which are based on the chord length $s$.}
%     \label{fig:mofat}
% \end{figure}

We assume that the clumps will be randomly orientated and randomly situated within the envelope. When we integrate out along our rays in Eq. \ref{eq:H} and interact with a clump, we will need to account for this randomness through the selection of a cord length $s$ (i.e., the distance to which the ray interacts, see Figure \ref{fig:mofat}). If a spheroid is placed in a uniform isotropic fluence of infinite straight lines, then the chord length distribution is described by \cite{kellerer_chord-length_1984} to be:
\begin{equation}\label{eq:chord_len}
    F(s) = 
    \frac{2s}{c_1}\left[c_2 + \frac{|1-\epsilon^2|^{0.5}}{4(s^{-2} -1)}\left\{ \left|\frac{1}{s^2}-1\right|^{0.5} \left(\frac{1}{s^3} +\frac{3}{2s} \right) +\frac{3}{2}\zeta(\frac{1}{s}) \right\}\right]
\end{equation} 
with
\begin{equation}
    \begin{aligned}
    \zeta(x) &= \left\{ \begin{aligned} 
    &cos^{-1}(x)~~~~~~ ~~~~~~~~~~~~~~~~~~~~ \mathrm{for} ~~~ 0 \leq x \leq 1 \\
    &ln(x+(x^2 -1)^{0.5})~~~~~~~~~~~ \mathrm{for} ~~~ 0 \leq x \leq 1 
    \end{aligned}\right.\\
    c_1 &= \frac{1}{2} + \frac{\epsilon^2}{2|1-\epsilon^2|^{0.5}}\zeta\left(\frac{1}{\epsilon}\right) ~~~~~~~\mathrm{and}~~~~~~
    c_2 = \frac{1}{4\epsilon^2} + \frac{3}{4}c_1,
    \end{aligned}
\end{equation}
where the first term, $c_2$, in the square bracket in Eq. \ref{eq:chord_len} applies only for $0 < x < \epsilon$, and the second term for $\epsilon < x < 1$ or $1 < x < \epsilon$. 
% It should be noted that this distribution is scaled to units of smaller axis length equal to 1 (see Figure 3 of \citealt{1984kellerer}).

\subsection{Implementation of clumps}\label{sec:appen_impl}
Following Figure \ref{fig:mofat}, the molecular region is defined by an inner radius, $r_{min}$, and an outer radius, $r_{max}$. The probability for a ray to interact with a clump in this region within some distance, $\Delta r$, is:

\begin{equation}\label{eq:prob_imp}
    P(r) = \frac{n(r)\Delta rS_c(\epsilon,r_c)}{4\pi r^2}.
\end{equation}
where $n(r)$ is the clump number density and is found through the relation:
\begin{equation}\label{eq:fc_Mco}
    \frac{M_{CO}}{f_c} = \int_{r_{1}}^{r_{1}+\Delta r}V_c(r,\epsilon)n(r)\rho(r) r^2 dr d\Omega
\end{equation}
where $M_{CO}$ is total molecular mass and $f_c$ is a density enhancement factor for the molecular density distribution, $\rho(r) = f_c \rho_{0}(r_{photo}/r)^n$, with $\rho_{0}$ being the scaled density based on $M_{CO}$ and the homogeneous case ($f_c=1$).

From Eqs. \ref{eq:semi_axes} to \ref{eq:fc_Mco}, we can now modify the solution of Eq. \ref{eq:I(p,z)}:
\begin{equation} \label{eq:I_clump}
    I^+(p,r,\nu) = I^+(p,r',\nu)e^{-\kappa(r)\rho(r)s^*} + \xi(r) S(r,\nu,\vec{v})
\end{equation}
with 
\begin{equation}\label{eq:xi}
     \begin{aligned}
        \xi(r) &= \left\{ \begin{aligned}     
           &2E_3(\tau^\dagger)  ~~~~~~~~~~~~~~~~~~~~~~\mathrm{for}~~~ f_c \ne 1\\
            &(1-e^{-\kappa\rho \Delta z})~~~~~~~~~~~~~~~\mathrm{for}~~~ f_c =1
            \end{aligned} \right.
     \end{aligned}
\end{equation}
where $s^*$ is a random chord length (Eq. \ref{eq:chord_len} scaled in relation to $r_a$ and $r_c$) and $\tau^{\dagger}$ is an effective optical depth found through the ensemble average of random chord half lengths in relation to the randomness of the clump position in our grid. $\tau^{\dagger}$ is akin to the average point of origin of the observed photons from a clump and is used in the solution of the plane-parallel atmosphere via the third exponential integral, $E_3$. 

% random chord half lengths treated with respect to the relative size of the clump compared to our grid size. $\tau^{\dagger}$ accounts for the random position where the emission may originate from in the clump. $E_3$ is the third exponential integral and accounts for self-absorption  It's emphasized that we use two separate probabilities when deciding if we include either the emission or absorption part of Eq. \ref{eq:I_clump}. \TBM{needs work after this} This process repeats as we integrate out along every $p_i$, where the reverse is also true for the $I^-$ direction. The final integral in Eq. \ref{eq:H} is then just sum of the flux like variables at the edge visible to the observer scaled to the the size of the supernovae. 
\section{Temperature and NLTE Corrections in MOFAT}\label{sec:temperature} 
Temperature and NLTE correction factors can be constrained through observed spectra. In principle, the underlying radiation field can be approximated through a combined analysis of spectral line profiles from the optical to the infrared. This information can then be used as boundary conditions for the initial temperature profile and can be incorporated into the rate equations. However, for certain epochs, this would require extensive SEDs, and may require a significant amount of computational time. Instead, we first assume that the temperature structure follows the photon density (to first order $T(^0v) \sim v^{-0.5}$) and then use the molecular features as the driving force in temperature structure and NLTE corrections. 

As described in Section \ref{sec:model}, we utilize $T(v)=T(v)+ C ~\Delta T(v)$ for our temperature correction scheme. To find the most optimal $T(v)$ for a given set of parameters, we iterate the shape and strength of $\Delta T(v)$, where each iteration influences the next until $C$ approaches unity (meaning that the flux level of our simulated spectrum is the same as the observation). Numerically, our solution takes on the form $T^{m+1}(v)=T^0(v)+ \Delta T^m(v)(1-f) + C^m\Delta T^{m-1}(v)f$, with $f$ being the fraction of the previous iteration $(m-1)$ to be used with the current iteration ($m$) in the upcoming iteration ($m+1$). As $C^m$ begins to approach unity $\Delta T(v)^{m-1}$ and $\Delta T(v)^{m}$ become close to equal. For SiO fitting, $T^0(v)$ is the temperature structure from the best CO fits.

\begin{figure}
    \centering
    \includegraphics[width=0.8\linewidth]{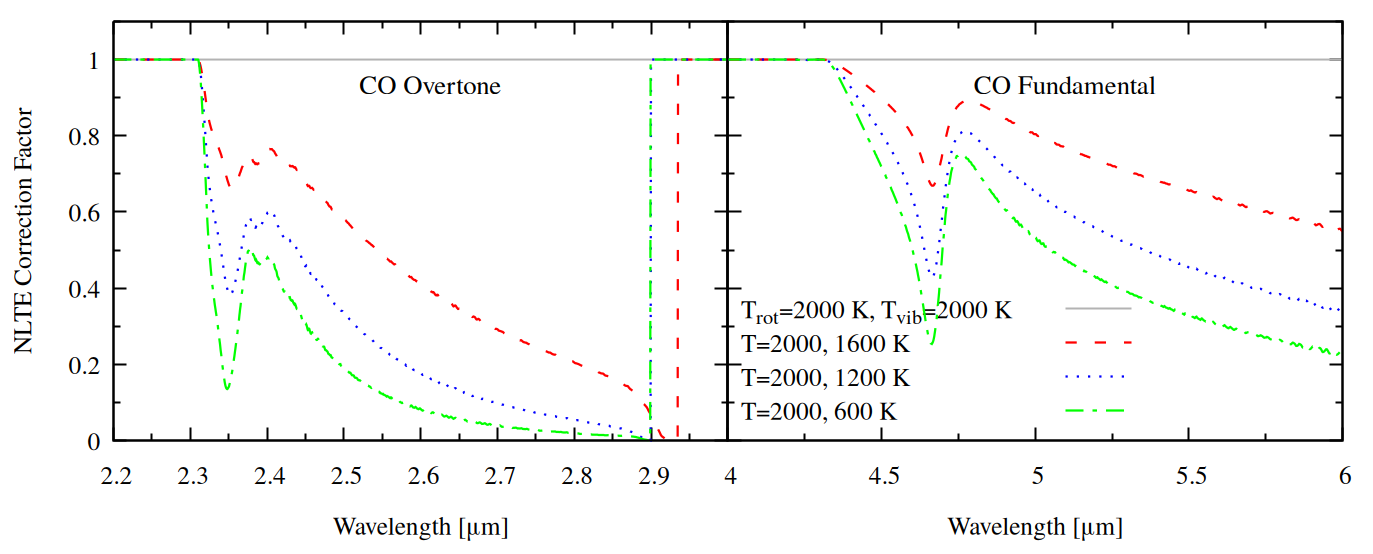}
    \caption{Demonstrating how the strength of decoupling between $T_{rot}$ and $T_{vib}$ can change the ratios between the modes of the CO overtone and fundamental bands for $T=2000$ K and no CO$^+$. We show $T_{vib}/T_{rot} = 1,~ 0.8,~0.6,~0.3$ and the NLTE correction factors when compared to LTE levels. }
    \label{fig:nlte}
\end{figure}

% For temperature corrections we utilize an analytic approach that uses the relation $L\sim B$:
% \begin{equation} \label{eq:temp_corr}
%     \begin{aligned}
%         \Delta T &\approx \left\{ \begin{aligned}
%             &\frac{d\Delta H}{d\tau_B}\frac{\pi}{4\sigma T^3} \\
%             & T\frac{\Delta L}{L}
%         \end{aligned} \right.
%     \end{aligned}
% \end{equation}
% where $\tau_B$ is the optical depth according to the Planck mean opacity for either the overtone or fundamental band. The selection of which temperature correction to use depends on the stability of the solution at a particular layer. The strength of $\Delta T$ is then driven by the observed overtone band as the inner boundary condition and the observed fundamental band as the outer boundary condition. 

The populations of the rotational levels in CO are more likely to stay in LTE compared to the vibrational levels. After about $\sim200$ days, the excitations keeping the vibrational levels in LTE are less than the radiative lifetimes, making the assumption of LTE progressively worse as time progresses \citep{liu_carbon_1992}. We do not consider an underlying radiation field or calculate electron density interactions when determining NLTE population levels; instead, we account for NLTE correction effects by assuming that the temperatures of the rotational and vibrational levels are decoupled $T_{rot} \ge  T_{vib}$. This assumption allows the population numbers to be readily computed \citep{gamache_extension_1992, rothman_hitemp_2010}, which determine the ratios between modes in the overtone and fundamental bands, depending on the degree of decoupling (see Figure \ref{fig:nlte}). The CO overtone is more sensitive to NLTE effects so we primarily use it to determine the most optimal $T_{vib}/T_{rot}$. After the $T_{vib}/T_{rot}$ has been found, it's held constant throughout the molecular forming region. For the reference model in Section \ref{sec:paramter_study_co}, we found $T_{vib}/T_{rot}=0.46$.

\section{Iteration Scheme in MOFAT}\label{sec:iteration}
MOFAT uses observations to derive an accurate description of the molecular region in an SN at a given time. It does this by finding an optimal set of parameters that best match the observations using a  $\chi^2_*$ nested-loop iteration scheme in a seven-dimensional parameter space. These parameters have been described in Section \ref{sec:model} and Table \ref{tab:model_values} with most of them being show in Figure \ref{fig:mofat}.
% The first five parameters are shown in Figure \ref{fig:mofat} and consist of the molecular region (defined by an inner velocity edge, $v_{min}$, with some width $\Delta v$) and the clumps (defined by density enhanced ($f_c\ge 1$) spheroids with semi-axes related to the radial dimension through $r_{R}$ and $\epsilon$). The final two parameters are the initial starting temperature $T_1$ (defined at $v_{min}$) and then the slope, $n$, of the molecular density distribution ($\rho\sim r^{-n}$). 

The basic structure of the iteration scheme used in MOFAT is given in Figure \ref{fig:mofat_iteration}. The outermost loop is set up to loop through various blocks. Here, we define the second outermost loop as a block. In a block, we make the selection of which parameters we want to iterate and then close those particular switches. The code then loops through each parameter, where each parameter variation also iterates temperature corrections ($T_{corr}$), total mass ($M_{CO}$), and NLTE strength until the $\chi_{*}^2$ statistic is minimized. This block can then be looped through multiple times, where between each block loop the selected parameters region is centered on the best fit value and subsequently shrunk by 10\%. For the next block, a different set of parameters can be chosen and the process repeats. 

% \begin{figure}
%     \centering
%     \includegraphics[width=0.65\linewidth]{fig12.png}
%     \caption{A schematic showing the iteration scheme used in MOFAT. The outermost loop loops through blocks, where each block closes particular switches for parameters it wants to test. Each selected parameter is then looped through, where temperature, NLTE, and total mass effects are iterated until the $\chi^2$ statistic reaches a convergence criterion or a maximum number of iterations. This block can then be looped through, where between loops the parameters region is centered on the best value and subsequently shrunk. This process repeats for any number of blocks. }
%     \label{fig:mofat_iteration}
% \end{figure}

\section{Sensitivity of parameters on S\lowercase{i}O features and evaluation of reference model} \label{sec:sensitivity_sio}

\begin{figure*}
    \centering
    \includegraphics[width=0.99\linewidth]{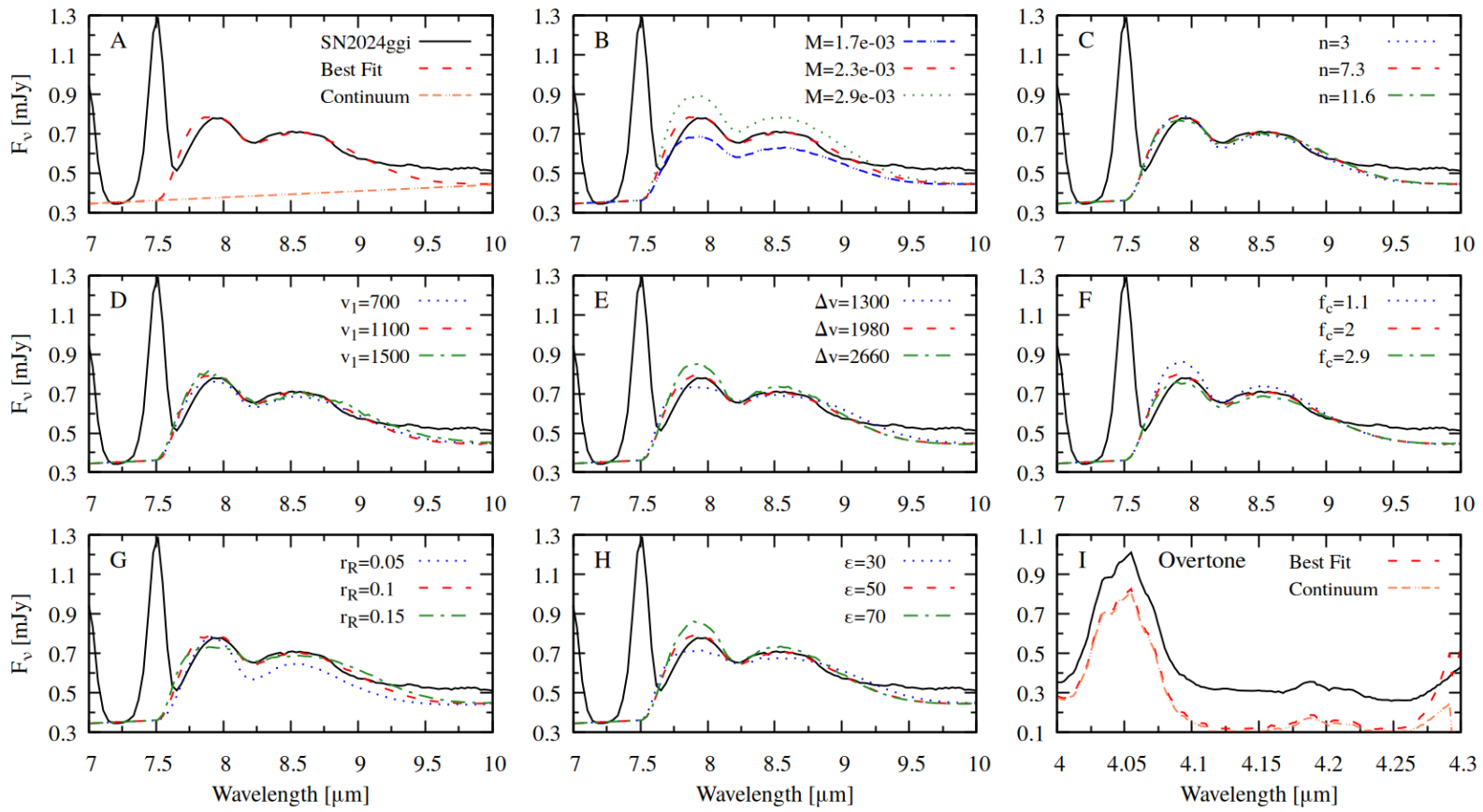}
    \caption{Demonstrating the sensitivity of each parameter on the SiO fundamental band of Model PP at day 385. Panel A) shows the best-fit solution and continuum. Each of the following panels keeps the optimized parameters the same but shows the effect of the: B) total molecular mass ($M$), C) density structure ($n$), D) inner velocity edge ($v_1$), E) width of molecular region ($\Delta v$), F) clump density contrast ($f_c$), G) clump size ($r_R$), and H) clump shape ($\epsilon$). Panel I) shows the lack of the SiO overtone band, where an upper mass limit was established by optimizing this feature. Figures \ref{fig:sio_chi} and \ref{fig:sio_chi_shape} show the $\chi^2_*$ convergences of the fits.}
    \label{fig:sio_spectrum}
\end{figure*}

\begin{figure*}
    \centering
    \includegraphics[width=0.99\linewidth]{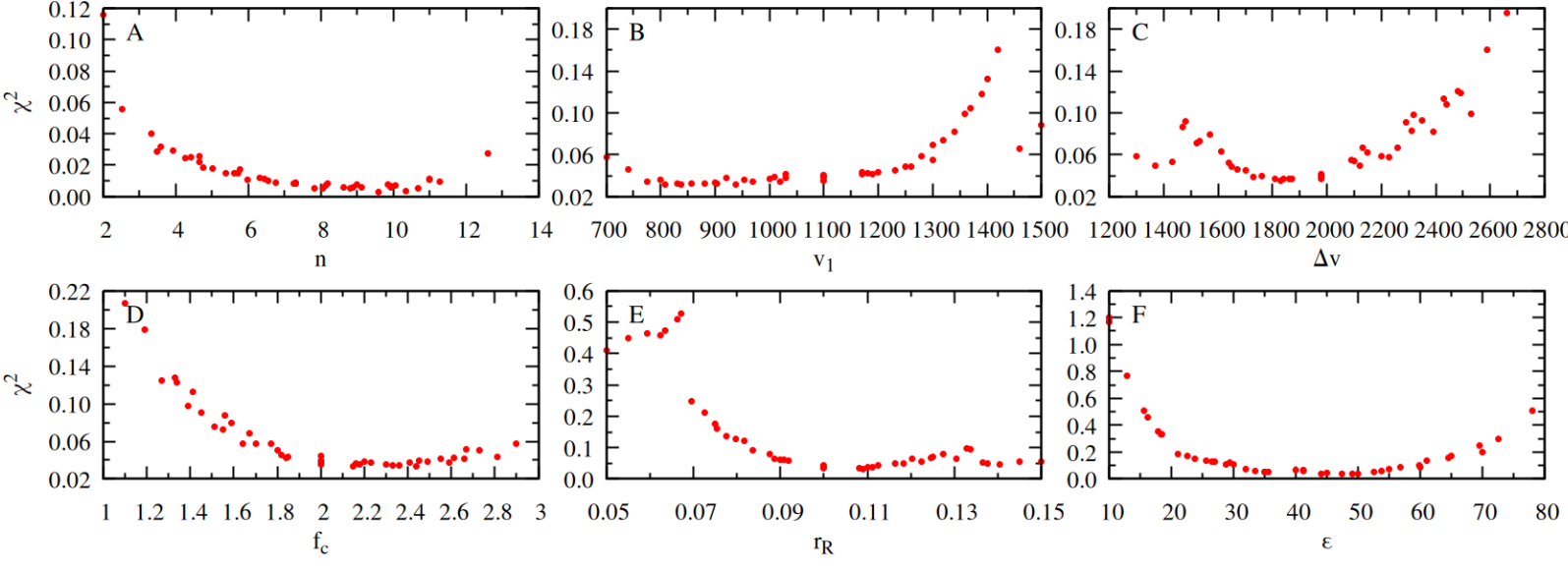}
    \caption{Demonstrating the $\chi ^2$ convergence for the fundamental band for each of the six primary parameters for Model PP at day 385. For each panel we keep the optimized parameters the same except for the parameter being varied. For panels A-F we only show $\chi^2$ of the fundamental band because the overtone is virtually non-existent. Panel A) shows density structure ($n$), B) inner velocity edge ($v_1$), C) width of the molecular region ($\Delta v$), D) clump density contrast ($f_c$), E) clump size ($r_R$), and F) clump shape ($\epsilon$). For discussion, see Section\ref{sec:sio_fits}.}
    \label{fig:sio_chi}
\end{figure*}

\begin{figure*}
    \centering
    \includegraphics[width=0.99\linewidth]{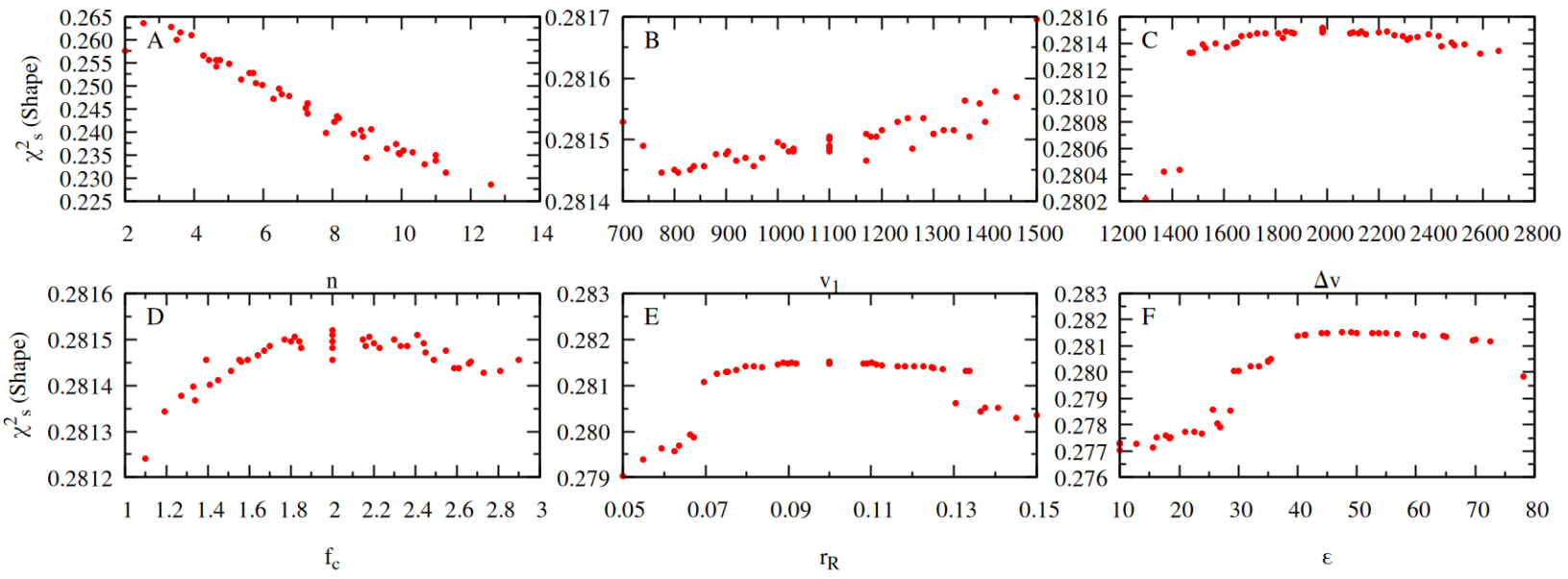}
    \caption{Same as Figure \ref{fig:sio_chi} except that we scale the absolute flux of our model to minimize the total flux difference with the observation and then calculate $\chi^2_s$. This measures the relative shape of the model compared to the observations. }
    \label{fig:sio_chi_shape}
\end{figure*}

In Figure \ref{fig:sio_spectrum}, we show the sensitivity of the parameters on the SiO fits for \ggi\ for Model PP at day 385. We do not show the effect of $T_1$ because this parameter is not iterated as it is synonymous with $v_1$ due to the $T(v)$ starting solution from the CO fits. The interpretation of the sensitivity of the parameters is similar to what was discussed in Section \ref{sec:paramter_study_co}. In Panel I, we show that our SiO overtone feature is limited to our best fit solution and can only be used to find an upper mass limit for a given continuum. From our solution, the continuum likely sits very close to the base of the $Br_\alpha$ and Mg features at $\sim4.05~\mu m$ and $\sim4.2~\mu m$, respectively. In Figures \ref{fig:sio_chi} and \ref{fig:sio_chi_shape}, we show the convergences of $\chi^2$ and $\chi^2_s$ of Model PP to \ggi\ at 385 days, respectively. Each parameter seems to converge to an optimal value, but because the convergences are very shallow and flat. With the additional effect that the $\chi^2_s$ stays flat in these parameter regions, we find it difficult to make any conclusion on the 3D morphologies of the SiO molecular region. 

\section{Correlation of Model Parameters}\label{sec:correlation}

The iterative scheme discussed in Appendix \ref{sec:iteration} efficiently narrows the large parameter ranges of our seven variables towards a solution with the lowest $\chi^2$ value. However, within such a high-dimensional parameter space, alternative solutions with comparable goodness of fit may exist. Due to the curse of dimensionality, randomly sampling the seven-dimensional parameter space to identify these solutions is computationally impractical. To explore the structure of the solution space and identify potential correlations between parameters, a large sample set is required.

\begin{figure*}
    \centering
    \includegraphics[width=\linewidth]{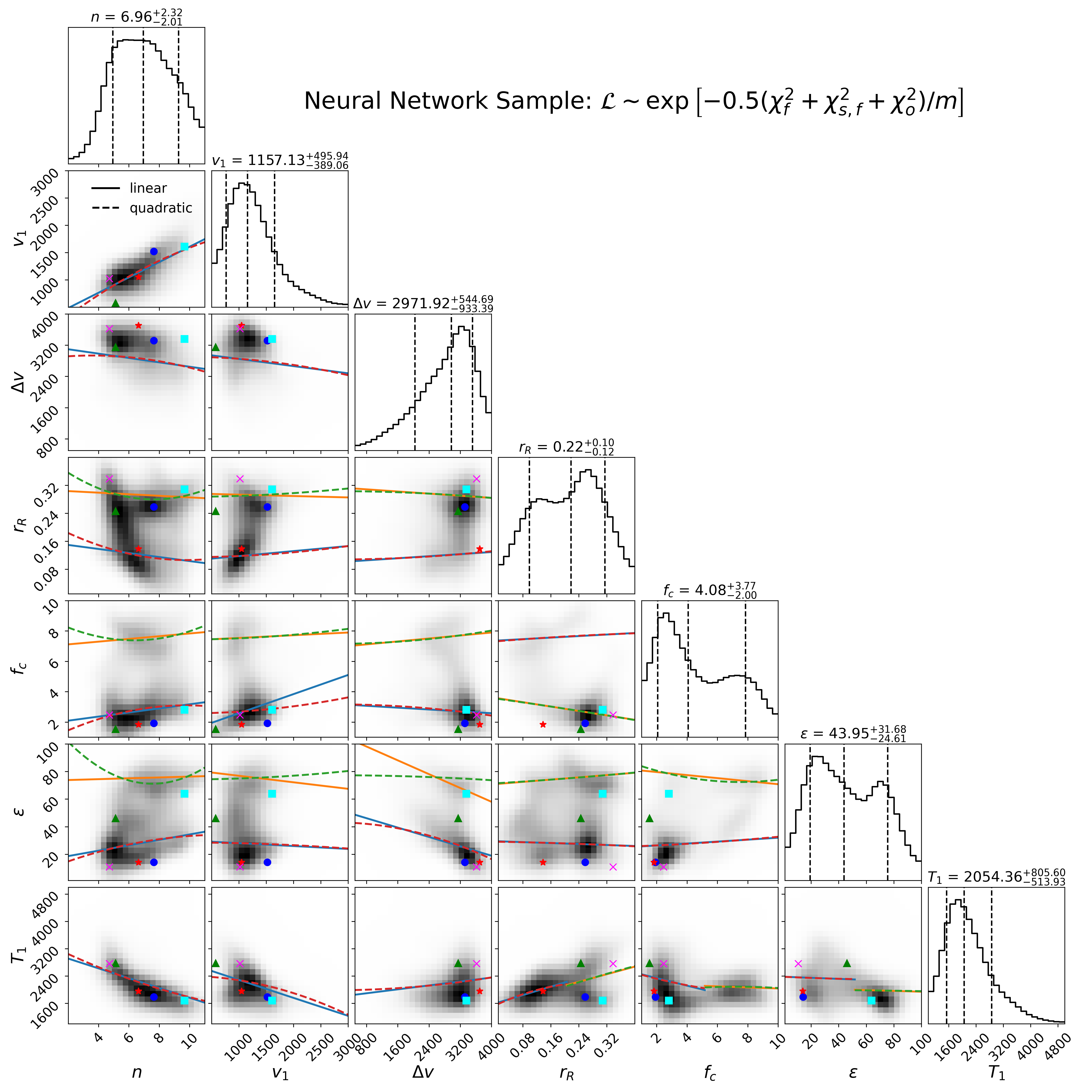}
    \caption{Corner plot showing the distribution of parameters for $10^5$ samples drawn from a normalizing flow model trained on $4000$ randomly sampled models in the seven-dimensional parameter space. The flow learns an approximation to the underlying likelihood-weighted parameter distribution, which is then sampled to generate a smooth representation of the posterior. Darker regions indicate higher posterior density. The likelihood weights ($w_i$) for the original models were computed from the combined statistic $\chi^2_{\alpha} = \chi_f^2 + \chi^2_{f,s}+\chi_o^2$, assuming Gaussian uncertainties, such that $w_i \propto \exp(-\chi^2_{\alpha,i}/2)$. For numerical stability, the minimum $\chi^2_{\alpha}$ value was subtracted prior to exponentiation. To mitigate extreme weight concentration, the original likelihoods were tempered by introducing a factor $1/m = 1/110$ in the exponential. Linear and quadratic fits are overlaid on each pairwise projection; the corresponding coefficients are listed in Table~\ref{tab:pairwise}. Parameter sets with $\chi_f^2$ (fundamental band) within 5\% of the reference model (red star) in Section~\ref{sec:paramter_study_co} are shown as points with distinct shapes and colors; their corresponding values are listed in Table~\ref{tab:other_best}. This plot was generated using the \texttt{corner.py} package \citep{corner}.}
    \label{fig:corner}
\end{figure*}

We therefore begin by randomly sampling the parameter space to construct a grid of 4000 models. Notably, only four of these models have $\chi^2$ values within 5\% of the reference model; their parameters are listed in Table \ref{tab:other_best}. Using this initial sample, we construct a likelihood grid based on the combined $\chi_*^2$ from the fundamental and overtone regions (see caption of Figure \ref{fig:corner}). We then use this likelihood grid to train a normalizing flow model, implemented via real-valued non-volume preserving (RealNVP) transformations \citep{2016dinh_realnvp, 2019paszke_pytorch}, to approximate the underlying parameter distribution and interpolate within the sparsely sampled space. Finally, we draw $10^5$ samples from the trained flow to generate a dense representation of the parameter space, enabling analysis of parameter correlations and degeneracies. We show the corner plot distribution of these samples in Figure \ref{fig:corner} and plot the reference model and models from Table \ref{tab:other_best} as points.

From Figure \ref{fig:corner}, we find that comparable solutions lie along a complex hypersurface within the parameter space. The parameters $r_R$, $f_c$, and $\epsilon$ exhibit bimodal distributions, indicating the presence of two (or more) distinct regions of solutions corresponding to different clump sizes and shapes. These regions may reflect different formation scales associated with neutrino-driven or Rayleigh–Taylor instabilities (see Section \ref{sec:motivation}). To analyze relationships involving these bimodal parameters, we apply the K-means clustering algorithm to partition pairwise sets with these parameters into two regions \citep{1979_hartigan_kmeans,2011pedregosa_scikit}. We then quantify all pairwise relationships using both the Pearson linear correlation coefficient \citep{1895pearson_lcor} and the distance correlation \citep{2008szekely_dcor}, with results listed in Table \ref{tab:pairwise}. From these metrics, we identify moderately strong linear relationships for the parameter pairs ($n,~v_1$), ($n,~T_1$), ($v_1,~T_1$), and ($v_1,~f_c$; region 1). The parameter $n$ governs the location of the decoupling region for the overtone band, dependent on $f_c$. In general, for lower $n$, the decoupling radius will shift to higher velocities. To avoid significant Doppler broadening effects, for both the overtone and fundamental band, $v_1$ will decrease to keep the decoupling radius at a lower velocity. This also introduces an inverse relationship with temperature: the overtone decoupling radius must coincide with temperatures near $T \sim 1900$ K, requiring higher values of $T_1$ at these lower values of $v_1$ for a decreasing $n$. Additionally, increasing $v_1$ reduces the optical depth of the fundamental band, leading to excess flux. To preserve flux levels, $f_c$ will increase, which restores sufficient opacity. The shape of the fundamental band will then be preserved by the clump size and shape. 

We also find moderate nonlinear relationships for ($T_1,~r_R$; region 1) and ($\epsilon,~\Delta v$; both regions), while most other parameter pairs exhibit weak dependencies. The relationship between $T_1$ and $r_R$ may reflect the evolutionary state of the CO clumps: at higher temperatures, the clumps are likely still forming or in an early destruction phase, corresponding to larger characteristic sizes. This is consistent with the trends reported in \cite{mera_jwst_2025}. The dependence between $\epsilon$ and $\Delta v$ may arise from pre-collapse mixing, which increases the spatial extent of the molecular formation region and moderates the development of Rayleigh–Taylor instabilities by smoothing chemical boundaries. This results in a balance between clump elongation (shape) and the size of the molecular formation region. 

To further characterize these trends, we perform linear and quadratic least-squares fits for all parameter pairs and overplot the resulting relations in Figure \ref{fig:corner}; the corresponding fit coefficients and $R^2$ values are listed in Table \ref{tab:pairwise}. Cubic functions were also tested, but yield negligible improvement in $R^2$. The $R^2$ values indicate that only a subset of parameter pairs exhibit moderate relationships. However, low $R^2$ values do not necessarily imply independence, but rather that the relationship is not well captured by the chosen functional form, particularly in the presence of nonlinear structure or multimodality. A more complete characterization of these relationships would require reconstruction of the underlying manifold(s) of viable solutions, using techniques similar to those described by \citet{2000_Tenenbaum_manifold}, \citet{2014constantine_subspace}, and \citet{2015transtrum_sloppy_models}. We defer this analysis to future work, where multiple CC~SNe can be jointly analyzed to further constrain the parameter space. Ideally, such datasets will include early-time observations ($\lesssim 250$ days) up to late-time observations ($\sim 500$ days), enabling direct constraints on the formation and evolution of clumps.

\begin{table*}[htb]
\centering
    \caption{Parameter values from the four points in Figure \ref{fig:corner} that have a $\chi^2$ value of the fundamental band within 5\% of the reference model in Table \ref{tab:model_values}. The \textit{CO Model} column tells us the color and shape of the point in Figure \ref{fig:corner}. The rest of the columns are the same as Table \ref{tab:model_values}.
    }
\begin{tabular}{c|c|c|c|c|c|c|c|c|c|c}
\hline
CO Model & Time [d] &Clumps&Mass [M$_\odot$] & T$_1$ [K] & v$_1$ [$km~s^{-1}$] & $\Delta$v [$km~s^{-1}$]& n    & $r_R$     & f$_c$   & $\epsilon$ \\ \hline 
Blue dot  &385&yes   & 1.84e-3 & 1787 & 1520 & 3320 & 7.65 & 2.58e-1 & 1.92 & 14     \\  
Green triangle &385 & yes & 7.62e-4 & 2776 & 574 & 3150 & 5.12 & 2.46e-1 & 1.53 & 46 \\
Cyan square & 385 & yes & 2.04e-3 & 1682 & 1610 & 3360 & 9.67 & 3.08e-1 & 2.81 & 64 \\
Magenta cross & 385 & yes & 8.90e-4 & 2762 & 1020 & 3620 & 4.72 & 3.38e-1 & 2.47 & 11 \\

\hline
\end{tabular}

    \label{tab:other_best}
\end{table*}

\begin{table*}[htb]
\label{tab:pairwise}
\centering
\caption{Summary of pairwise parameter relationships including linear correlation (Lcor), distance correlation (Dcor, measure of any non-linear relationship --- 0=no relationship, 1=near functional relationship), polynomial fit coefficients, and $R^2$. Polynomial coefficients correspond to fits of the form $y = a_1x + b_1$ or $y = a_2x^2 + b_2x + c$. $(R_1^2,~R_2^2)$ corresponds to the linear or quadratic fits, respectively. Note, low R2 values do not necessarily imply independenc. Region 1 corresponds to the red and blue lines in Figure \ref{fig:corner}, while region 2 corresponds to the orange and green lines. The linear fits are shown as solid blue or orange lines in Figure \ref{fig:corner}, while the quadratic fits are shown as dashed red or green lines. }
\begin{tabular}{c|c|c|c|c|c|c|c}
\hline
Parameter 1 & Parameter 2 & Region & Lcor & Dcor & $a_1,~b_1$ & $a_2,~b_2,~c$ & $R_1^2,~R_2^2$ \\
\hline

$n$ & $v_1$ &1& 0.512 & 0.559 & 140.4, 202.5 & -6.119, 229.8, -97.84 & 0.262, 0.266 \\ \cline{1-8}
$n$ & $T_1$ &1& -0.394 & 0.472 & -143.6, 3198 & 5.599, -225.4, 3473 & 0.155, 0.157 \\ \cline{1-8}
$n$ & $r_R$ &1& -0.201 & 0.238 & -5.764e-3, 0.1611 & 1.274e-3, -0.0250, 0.2286 & 0.040, 0.058 \\ \cline{3-8}
& &2 & -0.076 & 0.122 & -2.248e-3, 0.3078 & 2.393e-3, -0.0366, 0.4195 & 0.006, 0.055 \\ \cline{1-8}

$n$ & $f_c$ &1& 0.207 & 0.220 & 0.1350, 1.821 & -0.0273, 0.5329, 0.4964 & 0.043, 0.057 \\ \cline{3-8}
& &2 & 0.109 & 0.111 & 0.0897, 6.938 & 0.0449, -0.5702, 9.194 & 0.012, 0.038 \\ \cline{1-8}

$n$ & $\epsilon$ &1& 0.256 & 0.237 & 1.963, 14.73 & -0.1895, 4.584, 6.373 & 0.066, 0.070 \\ \cline{3-8}
& &2 & 0.023 & 0.210 & 0.3128, 73.19 & 0.9844, -14.88, 127.4 & 0.001, 0.051 \\ \cline{1-8}

$n$ & $\Delta v$ &1& -0.148 & 0.195 & -56.00, 3208 & -7.571, 54.64, 2836 & 0.022, 0.025 \\ \cline{1-8}

$v_1$ & $T_1$ &1& -0.396 & 0.397 & -0.5261, 2811 & -1.264e-4, -2.791e-8, 2404 & 0.157, 0.085 \\ \cline{1-8}

$v_1$ & $r_R$ &1& 0.115 & 0.217 & 1.435e-5, 0.1033 & 3.553e-9, 9.343e-13, 0.1143 & 0.013, 0.007 \\ \cline{3-8}
& &2 & -0.040 & 0.154 & -3.966e-6, 0.2971 & 2.642e-9, 5.413e-13, 0.2870 & 0.002, 0.007 \\ \cline{1-8}

$v_1$ & $f_c$ &1& 0.510 & 0.433 & 1.262e-3, 1.316 & 1.176e-7, 2.096e-11, 2.567 & 0.260, 0.019 \\ \cline{3-8}
& &2 & 0.062 & 0.071 & 1.777e-4, 7.363 & 7.788e-8, 2.069e-11, 7.439 & 0.004, 0.008 \\ \cline{1-8}

$v_1$ & $\epsilon$ &1& -0.076 & 0.137 & -2.102e-3, 30.16 & -4.886e-7, -1.367e-10, 28.54 & 0.006, 0.002 \\ \cline{3-8}
& &2 & -0.101 & 0.161 & -4.733e-3, 81.64 & 6.816e-7, 1.175e-10, 74.24 & 0.010, 0.002 \\ \cline{1-8}

$v_1$ & $\Delta v$ &1& -0.133 & 0.174 & -0.1841, 3032 & -5.252e-5, -1.160e-8, 2904 & 0.018, 0.014 \\ \cline{1-8}

$T_1$ & $r_R$ &1& 0.305 & 0.316 & 3079, 1593 & -8569, 4838, 1531 & 0.093, 0.098 \\ \cline{3-8}
& &2 & 0.212 & 0.209 & 2785, 1560 & -1566, 3955, 1358 & 0.045, 0.046 \\ \cline{1-8}

$T_1$ & $f_c$ &1& -0.187 & 0.200 & -111.3, 2562 & 6.220, -133.6, 2564 & 0.035, 0.036 \\ \cline{3-8}
& &2 & -0.042 & 0.061 & -15.61, 2190 & -4.999, 72.14, 1825 & 0.002, 0.005 \\ \cline{1-8}

$T_1$ & $\epsilon$ &1& -0.030 & 0.061 & -1.545, 2378 & 0.01156, -1.914, 2377 & 0.001, 0.001 \\ \cline{3-8}
& &2 & -0.049 & 0.191 & -1.209, 2059 & -3.784e-3, -5.017e-6, 1993 & 0.002, 0.009 \\ \cline{1-8}

$T_1$ & $\Delta v$ &1& 0.150 & 0.164 & 0.1436, 1779 & 2.340e-5, 4.118e-9, 1983 & 0.022, 0.016 \\ \cline{1-8}

$r_R$ & $f_c$ &1& 0.088 & 0.194 & 1.266, 7.360 & -1.317, 1.872, 7.310 & 0.008, 0.008 \\ \cline{3-8}
& &2 & -0.249 & 0.231 & -3.631, 3.602 & -0.9390, -3.193, 3.560 & 0.062, 0.062 \\ \cline{1-8}

$\epsilon$ & $\Delta v$ &1& -0.319 & 0.465 & -8.496e-3, 52.94 & -1.661e-6, -2.944e-10, 43.10 & 0.153, 0.178 \\ \cline{3-8}
& &2 & -0.386 & 0.465 & -0.01251, 108.1 & -2.235e-7, -3.830e-11, 77.28 & 0.149, 0.001 \\

\hline
\end{tabular}
\end{table*}

\clearpage

\label{lastpage}
\end{document}